\documentclass[longauth]{aa}
\usepackage{graphicx}
\usepackage{txfonts}

\usepackage{natbib}

\usepackage{diagbox}
\usepackage{rotating}
\usepackage{tikz}
\usepackage{multirow}
\usepackage{wasysym}
\usepackage{adjustbox}
\usepackage{afterpage}
\usepackage{placeins}
\usepackage{xcolor}
\usepackage{caption}
\usepackage[strict]{changepage}
\usepackage{lscape}
\usepackage{pdflscape}
\usepackage{soul}

\usepackage{hyperref}

\newcommand{\edit}[1]{\textcolor{black}{#1}}
\newcommand\boldfont[1]{\textcolor{black}{#1}}

\hypersetup{colorlinks=True,citecolor=blue}

\def\ms{\hbox{\,m\,s$^{-1}$}}         
\def\cms{\hbox{\,cm\,s$^{-1}$}}       
\def\m2s2{\hbox{\,m$^{2}$\,s$^{-2}$}} 
\def\kms{\hbox{\,km\,s$^{-1}$}}       

\def\logrhk{$\log$(R$^{\prime}_{HK}$) }

\begin{document}
   \title{A Decade of Solar High-Fidelity Spectroscopy and Precise Radial Velocities from HARPS-N}
   
   \titlerunning{A Decade of Solar High-Fidelity Spectroscopy and Precise Radial Velocities from HARPS-N}
   \authorrunning{X. Dumusque et al.}


\author{X. Dumusque \inst{1}
\and K. Al Moulla\inst{2}  \thanks{SNSF Postdoctoral Fellow}
\and M. Cretignier\inst{3}
\and N. Buchschacher \inst{1}
\and D. Segransan \inst{1}
\and D. F. Phillips \inst{4}
\and L. Affer \inst{5}
\and S. Aigrain \inst{2}
\and A. Anna John \inst{6}
\and A.~S.~Bonomo \inst{7}
\and V. Bourrier \inst{1}
\and L. A. Buchhave \inst{8}
\and A. Collier Cameron \inst{9}
\and H. M. Cegla \inst{10,11}
\and P. Cort\'es-Zuleta \inst{3}
\and R. Cosentino \inst{12}
\and J. Costes \inst{13}
\and M. Damasso \inst{7}
\and Z. L de Beurs \inst{14,15}
\and D. Ehrenreich \inst{1}
\and A. Ghedina \inst{12}
\and M. Gonzales \inst{12}
\and R. D. Haywood \inst{16}
\and B. Klein \inst{3}
\and B. S. Lakeland \inst{6}
\and N. Langellier \inst{17}
\and D. W. Latham  \inst{17}
\and A. Leleu \inst{1}
\and M. Lodi \inst{12}
\and M. Lopez-Morales \inst{18}
\and C. Lovis \inst{1}
\and L. Malavolta \inst{19,20}
\and J. Maldonado \inst{5}
\and G. Mantovan \inst{19,21}
\and A. F. Mat\'inez Fiorenzano \inst{12}
\and G. Micela \inst{5}
\and T. Milbourne \inst{22}
\and E. Molinari \inst{23}
\and A. Mortier \inst{6}
\and L. Naponiello \inst{7}
\and B. A. Nicholson \inst{24}
\and N. K. O'Sullivan \inst{2}
\and F. Pepe \inst{1}
\and M. Pinamonti \inst{7}
\and G. Piotto \inst{19}
\and F. Rescigno \inst{6}
\and K. Rice \inst{25}
\and S. Dimitar \inst{17}
\and A. M. Silva \inst{2}
\and A. Sozzetti \inst{7}
\and M. Stalport \inst{26,27}
\and S. Tavella \inst{1,28}
\and S. Udry \inst{1}
\and A. Vanderburg \inst{14}
\and S. Vissapragada \inst{17}
\and C. A. Watson \inst{13}
}

   \institute{Astronomy Department of the University of Geneva, 51 ch. des Maillettes, 1290 Versoix, Switzerland\\
              \email{xavier.dumusque@unige.ch}
    \and Instituto de Astrofísica e Ciência
    s do Espaço, Universidade do Porto, CAUP, Rua das Estrelas, 4150-762 Porto, Portugal
    \and Denys Wilkinson Building, Department of Physics, University of Oxford, OX1 3RH, UK
    \and MIT Lincoln Laboratory, Lexington, MA 02421, USA
    \and INAF – Osservatorio Astronomico di Palermo, Piazza del Parlamento 1, 90134 Palermo, Italy
    \and School of Physics \& Astronomy, University of Birmingham, Edgbaston, Birmingham B15 2TT, UK
    \and INAF - Osservatorio Astrofisico di Torino, via Osservatorio 20, 10025 Pino Torinese, Italy
    \and DTU Space, National Space Institute, Technical University of Denmark, Elektrovej 328, DK-2800 Kgs. Lyngby, Denmark
    \and SUPA School of Physics and Astronomy, University of St Andrews, North Haugh, St Andrews KY16 9SS, UK
    \and Physics Department, University of Warwick, Gibbet Hill Road, Coventry CV4 7AL, UK
    \and Centre for Exoplanets and Habitability, University of Warwick, Gibbet Hill Road, Coventry CV4 7AL, UK
    \and Fundaci\'on Galileo Galilei-INAF, Rambla Jos\'e Ana Fernandez P\'erez 7, E-38712 Bre\~na Baja, Tenerife, Spain
    \and Astrophysics Research Centre, School of Mathematics and Physics, Queen’s University Belfast, Belfast, BT7 1NN, UK
    \and Department of Earth, Atmospheric and Planetary Sciences, Massachusetts Institute of Technology, Cambridge, MA 02139, USA
    \and Kavli Institute for Astrophysics and Space Research, Massachusetts Institute of Technology, Cambridge, MA 02139, USA
    \and Astrophysics Group, University of Exeter, Exeter EX4 2QL, UK
    \and Center for Astrophysics | Harvard \& Smithsonian, 60 Garden Street, Cambridge, MA 02138, USA
    \and Space Telescope Science Institute, 3700 San Martin Drive, Baltimore, MD 21218, USA
    \and INAF, Astronomical Observatory of Padua, Vicolo dell’Osservatorio 5, I-35122 Padua, Italy
    \and Department of Physics and Astronomy ‘Galileo Galilei’, University of Padova, Vicolo dell’Osservatorio 3, I-35122 Padova, Italy
    \and Centro di Ateneo di Studi e Attività Spaziali ‘G. Colombo’, Università degli Studi di Padova, Via Venezia 15, IT-35131, Padova, Italy
    \and Moses Brown School, Providence, RI 02906, USA
    \and INAF - Osservatorio Astronomico di Cagliari, via della Scienza 5, 09047, Selargius, Italy
    \and Centre for Astrophysics, University of Southern Queensland, Toowoomba, QLD 4350, Australia
    \and SUPA, Institute for Astronomy, University of Edinburgh, The Royal Observatory, Blackford Hill, Edinburgh EH9 3HJ, UK
    \and Space sciences, Technologies and Astrophysics Research (STAR)
    Institute, Université de Liège, Allée du 6 Août 19C, 4000 Liège,
    Belgium
    \and Astrobiology Research Unit, Université de Liège, Allée du 6 Août 19C, 4000 Liège, Belgium
    \and European Southern Observatory, Av. Alonso de Cordova, 3107, Vitacura, Santiago de Chile, Chile
}

   \date{Received XXX ; accepted XXX}

 
  \abstract
   {The HARPS-N solar telescope has been observing the Sun every possible day since the summer of 2015. We recently released 10 years of those data, which are available at \url{https://dace.unige.ch/openData/?record=10.82180/dace-h4s8lp7c}.}
  {The goal of this manuscript is to present the different optimisations made to the ESPRESSO data reduction software used to extract the published HARPS-N solar spectra, to describe data curation, and to perform some analyses that demonstrate the extreme RV precision of those data.}
   {By analysing all the HARPS-N wavelength solutions over 13 years, we bring to light instrumental systematics at the 1\ms\,level. We mitigate those systematics by curating the thorium line list used to derive the wavelength solution, and by applying a correction to the drift of thorium lines induced by the aging of Thorium-Argon hollow cathode lamps. After optimisation, we demonstrate a peak-to-peak precision on the HARPS-N wavelength solution better than 0.75\ms over 13 years. We then carefully curate the decade of HARPS-N re-reduced solar observations by rejecting 30\% of the data affected either by clouds, \boldfont{bad atmospheric conditions} or well-understood instrumental systematics. 
    Finally, we correct the curated data for spurious sub-\ms RV effects caused by erroneous instrumental drift measurements and by changes in the spectral blaze function over time.}
   {After curation and correction, a total of 109,466 HARPS-N solar spectra and respective RVs over a decade are available. The median photon-noise precision of the RV data is 0.28\ms and, on daily timescales, the median RV rms is 0.49\ms, similar to the level imposed by stellar granulation signals. On 10-year timescales, the large RV rms of 2.95\ms results from the RV signature of the Sun's magnetic cycle. When modeling this long-term effect using the Bremen Composite Magnesium II activity index, we demonstrate a long-term RV precision of 0.41\ms. We also analysed contemporaneous HARPS-N and NEID solar RVs and found the data from both instruments to be of similar quality and precision. However, an analysis of the RV difference between these two RV datasets over the three available years gives a surprisingly large RV rms of 1.3\ms. This variation is dominated by an unexplained trend that could be caused by a different sensitivity to stellar activity of the two datasets. Once this trend is modeled, the overall RV rms for three years reaches 0.79\ms, and the RV rms during the low activity phase decreases to 0.6\ms, compatible with what is expected from supergranulation.}
   {This decade of high-cadence HARPS-N solar observations with short- and long-term precision below 1\ms represents a crucial dataset to further understand stellar activity signals in solar-type stars , and to advance other science cases requiring such an extreme precision.}


   \maketitle

%
%
%
%


\section{Introduction}

The detection of an Earth-mass planet orbiting in the habitable zone of a solar-type star is still out of reach even for the most advanced, precision radial-velocity (RV) instruments currently available.
\edit{The RV signal induced by such a planet is tiny, with an amplitude of 9 \cms varying over hundreds of days.
Instrumental systematics, which can be many times larger than this signal, and stellar variability, which can be larger by orders of magnitude, can completely obscure the signal from an Earth-mass planet.}

Although instrumental calibrations should be able to probe and correct for instrumental systematics, stellar light is different in nature than calibration sources and the same instrumental shift measured using a calibration or a stellar spectrum will differ. This is noticeable if line-spread function (LSF) variations over the detector and over time are not modeled properly. Such effects are starting to be modeled in extracted spectra thanks to laser frequency comb calibration sources \citep[][]{Schmidt:2024aa, Hirano:2020aa}, but critical next steps are to optimise the modelisation and perform it at the raw frame level \citep[e.g.][]{Bolton:2010aa,Piskunov:2021aa}. In addition, understanding and mitigating stellar variability, which is thought to be the dominant barrier to detecting Earth-like planets, requires a large number of high-quality stellar spectra \citep[e.g.][]{Crass:2021aa}.

These arguments motivate daily observations of very bright stars with the best possible precision to track stellar variability and instrumental signals on the long term. Among all the bright stars available, the Sun has the advantage of i) being observable during the day, therefore not conflicting with night-time observations, ii) being very bright, therefore producing high signal-to-noise ratio (S/N) spectra, iii) having a solar system for which the gravitational pull of the orbiting planets is known with a precision better than one\cms, and iv) presenting disc-resolved observations from many ground and space-based telescopes to understand the nature of stellar variability seen at the RV or spectral level.

The first solar telescope built for the purpose of obtaining extreme precision radial velocities (EPRV) started operations in July 2015 using the HARPS-N spectrograph at the Telescopio Nazionale Galileo in Roque de Los Muchachos Observatory, La Palma island, Spain \citep[][]{Dumusque-2015b, Phillips:2016aa}. 
Its main feature is the use of an integrating sphere that \edit{mitigates guiding issues and injects uniform}, integrated light of the solar disc into a fibre that is then routed to the calibration unit of HARPS-N. 
\edit{Previous Sun-as-a-star RV observations consisted of sunlight being directly focused onto an optical fibre \citep{Lemke-2016}.}
\edit{Unfortunately, this approach showed guiding limitations which induced RV variations at the \ms level.}
Following this first experiment on HARPS-N, most of the other extreme precision spectrographs have been or will be equipped with a similar solar feed, all using the solution of an integrating sphere. A non-exhaustive list of existing or planned solar telescopes includes: HELIOS feeding sunlight into HARPS and more recently NIRPS \citep[][started operations in 2018]{AlMoulla:2023aa}, the LOST telescope feeding sunlight into EXPRES \citep[][started operation in 2020]{Llama:2022aa, Llama:2024aa}, the NEID solar telescope \citep[][started operations in 2021]{Lin:2022aa}, LOCNES the solar telescope for GIANO \citep[][started operations in 2023]{Claudi:2018aa}, SoCal the KPF solar telescope \citep[][started operations in 2023]{Rubenzahl:2023aa} and in the near future the PoET solar telescope feeding \edit{both disc-integrated and} disc-resolved images of the Sun into ESPRESSO \citep[][]{Santos:2023aa,Santos:2025aa} and the ABORAS solar telescope capable of observing the Sun in \edit{Stokes-V} spectro-polarimetry using HARPS-3 \citep[][]{Farret-Jentink:2022aa}. 
We note that in \citet{Zhao:2023ab}, the authors performed a detailed comparison of solar data obtained over the same month from HARPS-N, HARPS, NEID and EXPRES and find excellent agreement at the\ms level between all the datasets.

\edit{The first three years of HARPS-N solar data were published by \citet{Collier-Cameron:2019aa}.}
This first data release, providing RVs and cross-correlation function (CCF) profile parameters, was followed by a second release of the same time span but providing better short-term and long-term RV precisions, in addition to wavelength calibrated extracted spectra and CCFs \citep[][]{Dumusque:2021aa}. 
We note that the data from NEID and KPF are directly made public after the observations through the NASA Exoplanet Science Institute NEID Solar Archive \footnote{\url{https://neid.ipac.caltech.edu/search_solar.php}.} and the Keck Observatory Archive \footnote{\url{http://koa.ipac.caltech.edu/cgi-bin/KOA/nph-KOAlogin}.}, respectively. HELIOS raw data are made available through the ESO archive \footnote{\url{https://archive.eso.org/eso/eso_archive_main.html}, select \emph{Program ID 1102.D-0954} and \emph{Category} SCIENCE.}
%

\edit{High-fidelity spectra and extreme-precision RVs from the HARPS-N solar telescope, both from the previous releases described by \citet{Collier-Cameron:2019aa, Dumusque:2021aa} and the data described in this paper, have been used in many different studies to better understand and mitigate the effects of stellar variability, from both the HARPS-N consortium and other teams internationally.}
To better characterise stellar magnetic activity and its effect on RVs, several studies looked at different types of activity indicators and how they correlate with the RVs \citep[][]{Maldonado:2019aa, Milbourne:2019aa,  Thompson:2020aa, Milbourne:2021aa, Haywood:2022aa,
Sen:2023aa, Lienhard:2023aa,Cretignier:2024aa,Dravins:2024aa}. The data were also used to demonstrate that the stellar activity seen in RVs depends on the formation depth of the spectral lines \citep[][]{Cretignier:2020aa,AlMoulla:2022aa,Rescigno:2025aa} as originally discussed in \citet{Meunier:2017aa}. Comparison between realistic magnetic activity solar simulations and HARPS-N spectral and RV data also bring a much deeper understanding of the origin and link between activity and observed RV signal \citep[][]{Meunier:2024aa, Zhao:2025aa}. The HARPS-N solar data were also used to probe stellar signals induced by granulation \edit{and supergranulation} \citep[][]{AlMoulla:2023aa, Lakeland:2024aa, OSullivan:2025aa} and also to demonstrate that flares have a negligible contribution to RV variations in quiet solar-type stars \citep[][]{Pietrow:2024aa}.

HARPS-N solar data have also been used to develop new mitigation techniques for stellar signals. From a \edit{purely} data-driven approach, the SCALPELS \citep[][]{Collier-Cameron:2021aa} and YARARA \citep[][]{Cretignier:2021aa,Cretignier:2022aa,Cretignier:2023ab} frameworks were developed and optimised thanks to the HARPS-N solar data. \citet{Klein:2024aa} used these data to demonstrate how a SCALPELs approach linked to a multidimensional GP could be used to efficiently mitigate stellar activity and recover a planetary signal with amplitude as small as 0.4\ms on long-period orbits. 
\edit{\citet{Zhao:2022ab} use these data to demonstrate the efficiency of decomposing the CCF in Fourier space to separate true Doppler shifts from other perturbing systematic components. }
Those same data were also used to develop neural-network based approaches to mitigate stellar activity, as published in \citet{Beurs:2022aa}, \citet{Zhao:2024cc} and \citet{Colwell:2025aa}, where the authors demonstrate that planets as small as 20 cm/s on long-period up to 500 days could be detected. \citet{Klein:2025aa} demonstrated that\edit{, owing} to the high-fidelity of the solar spectra, it is possible to perform Doppler imaging even in the case of the very slow solar rotation and use the information to mitigate stellar activity signals.

\edit{Another use of the HARPS-N solar data, in addition to the stated science goals of studying the effects of stellar variability, has been for probing instrumental systematics over time. }
If spectrograph calibrations \edit{do not perfectly account for} instrument instabilities, any stellar data will be affected. These effects might be difficult to probe on stars due to the uneven and sparse sampling inherent to night-time observations at a \edit{telescope with high demands on observation time.}
\edit{Fortunately, the near-continuous nature of the solar observations allows us to easily spot these effects.}
A solar telescope can therefore be used as a calibrator for high-resolution EPRV spectrographs aiming at sub-m/s precision. 
\edit{Although this avenue of calibration techniques is ongoing work and yet to be fully explored, results presented in \citet{Dumusque:2021aa}, \citet{Cretignier:2021aa}, \citet{Ford:2024aa} and in this paper show the potential of solar data for better characterization of EPRV spectrographs.}

The goal of this study is to present the data reduction of a decade of HARPS-N solar observations, and demonstrate the RV precision of the obtained data. In Sect.~\ref{sect:DRS_optimisation} we describe the optimization of the ESPRESSO Data Reduction Software\footnote{The ESPRESSO data reduction software is publicly available on the ESO website, \url{https://www.eso.org/sci/software/pipelines/espresso/espresso-pipe-recipes.html}.} (DRS) to ensure the best possible short-term and long-term precisions for the solar data despite using hollow cathode lamps for wavelength calibration. We then discuss in Sect.~\ref{RV_data_set} the solar data themselves, and the curation performed to reject imprecise measurements. We also describe in this section two post-processing corrections of instrumental systematics affecting our solar measurements. In Sect.~\ref{results}, we demonstrate the RV precision of this decade of HARPS-N solar observations and compare those data to data taken with NEID solar telescope. We finally discuss further prospects and conclude in Sect.~\ref{Conclusion}.

\section{Data reduction software optimisation} \label{sect:DRS_optimisation}

The ESPRESSO DRS has now been adapted to reduce HARPS-N raw frames, and therefore delivers extracted spectra and RVs \edit{from the raw} instrument data. \citet[][]{Dumusque:2021aa} describe this implementation and details the benefit of this new data reduction compared to the original HARPS-N DRS. One of the major improvements is the use of a new algorithm to extract a wavelength solution from a thorium-argon (TH-AR) hollow-cathode (HC) lamp spectrum, which significantly reduces day-to-day scatter from $\sim$0.8 to $\sim$0.5\ms. The previous high scatter was induced by instabilities in the polynomial fit performed on the TH lines' wavelength and position to extract a wavelength for each extracted pixel (commonly known as wavelength solution). 

The HARPS-N solar data presented in \citet[][]{Dumusque:2021aa} correspond to version 2.2.3 of the ESPRESSO DRS, while we present here solar data reduced with version 3.2.5. Although most of the differences between those DRS versions are related to the optimisation of the ESPRESSO DRS for the NIRPS near-infrared instrument, some optimisations are related to HARPS-N. Among those, the ESPRESSO DRS implements a correction for the drift of TH lines related to the aging of the TH-AR lamp and its replacement in May 2020 (see Sect.~\ref{sect:th_drift}). In addition, with a new TH-AR lamp in use, and a larger number of spectrograph calibrations compared to what was available at the time of \citet[][]{Dumusque:2021aa}, we revisited the TH line list used for the wavelength solution and optimised the corresponding algorithm. Overall, version 3.2.5 of the ESPRESSO DRS significantly improves the long-term stability of the RVs while keeping a similar short-term precision as version 2.2.3. We note that the ESPRESSO DRS version 3.2.5 also includes a correction for telluric absorption spectral lines, based on the framework described in \citet{Allart:2022aa}, however, the solar data presented here do not benefit from that correction since it is not yet validated for HARPS-N spectra. Telluric contamination in the visible wavelength range of HARPS-N is, however, much smaller than in the near-infrared, like for example in NIRPS. In addition, since Earth is orbiting around the Solar system barycentre (SSB), the main contribution to the barycentric Earth RV is not the 30 km/s velocity of the Earth along its orbit, since it is always nearly perpendicular to the SSB direction, but to the velocity induced by Earth's rotation ($\sim500$\ms) and the tiny perturbation induced by the Earth orbit eccentricity (e = 0.017). The barycentric Earth RV is thus contained within $\pm0.9$\kms, which implies that telluric lines do not shift by more than 2 pixels relative to stellar lines\footnote{the average pixel size on HARPS-N is 820\ms.}, thus strongly reducing the impact of telluric contamination. Although we note that in the ESPRESSO G2 mask used to obtain the RVs through cross-correlation, stellar lines falling close to the main telluric features have been discarded, telluric lines might still induce some RV bias in techniques such as template matching \citep[][]{Silva:2025aa}.

\subsection{Compensation for the offset of Thorium lines introduced by lamp aging}\label{sect:th_drift}

\begin{figure}
        \center
	\includegraphics[width=9cm]{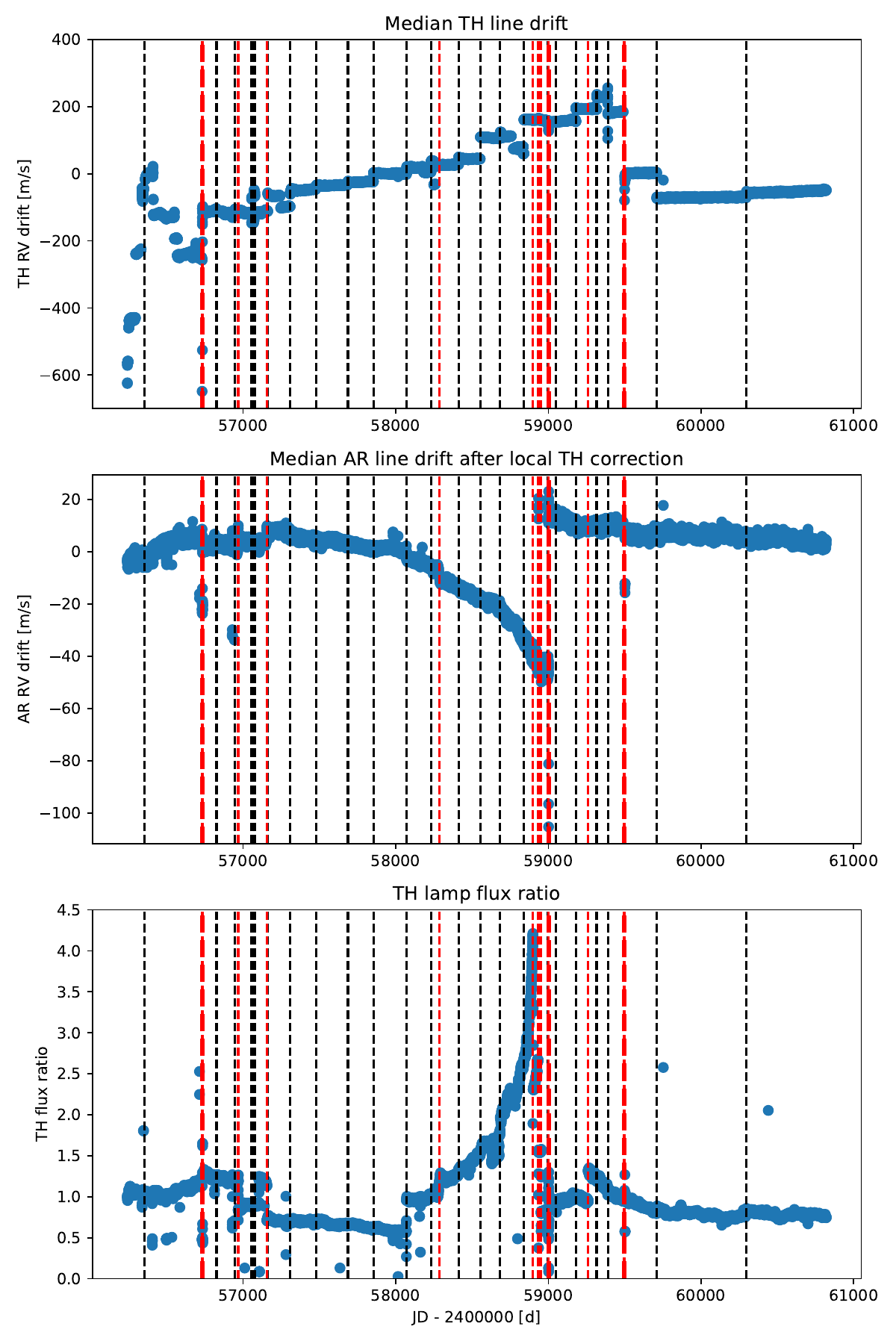}
	\caption{\emph{Top:} Median drift of all TH lines as a function of time. This tracks the drift of HARPS-N over time, which stays below \edit{1} \kms despite several drastic instrumental interventions. \emph{Middle:} Median behaviour of the AR lines as a function of time once the drift of TH lines \edit{has} been corrected for\edit{.} 
    \edit{This correction is applied} locally on different regions of the detector to be less sensitive to LSF variations (see text). \emph{Bottom:} TH flux ratio, i.e.
    integrated flux of the TH lamp with respect to a reference, as a function of time. We see that between JD = 2458500 and 2459002 (16 January 2019 to 2 June 2020) the TH flux ratio reached very high values, which will have an impact on wavelength solution derivation. The thick red dashed vertical lines correspond, chronologically to the change of the HARPS-N focus, the change of the TH-AR HC lamp used for wavelength solution and the change of the detector cryostat.
    The thin red vertical dashed lines correspond to sudden changes in the flux of the TH-AR HC lamp (see bottom panel), and the black vertical dashed lines correspond to detector warm-ups. More details about those events are given in Table~\ref{app:intervention}} 
	\label{fig:th_ar_drift}
\end{figure}

TH-AR HC lamps have been used as a standard for precise and accurate wavelength solutions of high-resolution spectrographs for decades \citep[e.g.][]{Kerber:2007aa}.
\edit{Although} many studies \edit{over the years} have improved our understanding of the TH-AR HC lamp emission spectrum \citep[e.g.][]{Palmer:1983aa, Lovis-2007b, Redman:2014aa}, very little is known about the stability of the output spectrum with lamp aging at the \ms level. This is challenging to measure, and cannot be obtained by looking at the stability of wavelength solutions as a function of time. This is because the HC lamp emission spectrum is used as reference to calibrate a spectrograph that, even if extremely stabilised, nevertheless drifts by hundreds of \ms over several years (see top panel of Fig.~\ref{fig:th_ar_drift} for HARPS-N). 
\edit{One way to investigate the effect of the lamp aging is to take a monthly cross-calibration between a wavelength solution derived from the HC lamp used for daily calibrations, and a second HC lamp that is not used daily, and thus ages more slowly.}
Another possible way is to cross-calibrate the HC lamp with a laser-frequency comb (LFC), which is \edit{designed for extreme long-term stability}. 
We note that some studies try to bypass this understanding of how HC lamps age\edit{, and the effect this has} on derived RVs, by directly analysing the variability of \emph{RV standard stars} to correct for instrumental systematics \citep[e.g.][]{Trifonov-2020}. However, this is challenging and prone to error in the context of EPRV, as we know that at the \ms level, the RV of all stars are variable due to stellar signals, mainly in this case the long-term RV effect induced by magnetic cycles \citep[e.g.][]{Meunier-2010a}.

No useful cross-calibration between the TH-AR HC lamp used for daily wavelength solution and a lamp of reference was performed on HARPS-N. Although an experimental LFC \citep[][]{Phillips-2012a} was available, the frequency of LFC spectra measurements and the change in the LFC spectrum due to interventions over time, make it difficult to study how the spectrum of the used TH-AR HC lamp behaved over time. At the end of May 2020, the TH-AR HC lamp used since the beginning of HARPS-N operation in April 2012 had to be replaced by a spare. This single event is equivalent to a cross-calibration between a very old and a new lamp, if we assume that the spectrograph did not significantly drift during the lamp change intervention. In Appendix~\ref{app:th_ageing}, we analyse this lamp change on HARPS-N, in addition to two lamp changes on HARPS that happened since HARPS was equipped with octogonal fibres \citep{LoCurto:2015aa}. 
\edit{We find that, while an RV offset is observed for both the TH and AR lines, the observed offset for the AR lines is significantly higher than for the TH lines.}
In particular, the offset ratio between AR ground state lines (AR1) and TH lines seems consistent between the three lamp changes analysed, with a mean value of $\mathrm{offset}_{AR1}/\mathrm{offset}_{TH} = 20.95$. Now, if we assume that this offset ratio is constant throughout the lifetime of a lamp, we can measure the offset of TH lines due to lamp aging over time by measuring the drift of AR1 lines. To do so, we compare the measured wavelength of AR1, given by the spectrograph wavelength solution\edit{,} to their laboratory wavelength (or any reference wavelength at a given time). As the wavelength solution is provided by TH lines, the value that we measure is not the drift of AR1 lines, but the drift difference $\mathrm{offset}_{AR1} - \mathrm{offset}_{TH}$. As demonstrated in Appendix~\ref{app:th_ageing}, the TH lines offset can be expressed as:
\begin{equation}
\mathrm{offset}_{TH} = \frac{\mathrm{offset}_{AR1} - \mathrm{offset}_{TH}}{\mathrm{offset}_{AR1}/\mathrm{offset}_{TH}-1} = \frac{\mathrm{offset}_{AR1} - \mathrm{offset}_{TH}}{19.95}
\end{equation}
The behaviour of $\mathrm{offset}_{AR1} - \mathrm{offset}_{TH}$ as a function of time can be seen in the middle panel of Fig.~\ref{fig:th_ar_drift}. As already mentioned, this correction is based on the hypothesis that the $\mathrm{offset}_{AR1}/\mathrm{offset}_{TH}$ ratio is constant throughout the lifetime of a lamp, which we do not demonstrate here. 
\edit{When we fit this ratio to the solar data, we find a value of $\mathrm{offset}_{AR1}/\mathrm{offset}_{TH}-1 = 15.83$, slightly lower than the 19.95 value obtained from studying offsets at the lamp changes in Appendix~\ref{app:th_ageing}. See Sect.~\ref{sect:opti_th_drift} for more details.}
Correcting for the drift of TH lines over time is a challenging task, and we advise any team still relying on HC lamp for wavelength calibration to cross-calibrate the lamp they use with a reference that is only used once a month to track effects due to lamp aging.

%
%
%

\subsection{Thorium line curation for optimising long-term wavelength solution stability}\label{sect:wave_sol}

The long-term stability of the wavelength solution relies mainly on the long-term stability of the TH lines used as input to any wavelength solution algorithm. For this new solar data release, we revisited the work done in Appendix C of \citet{Dumusque:2021aa} regarding TH line selection. The details about this new selection of TH lines are presented in Appendix~\ref{app:th_curation}.

After curation, we are left with 2020 TH lines well distributed on the HARPS-N detector (see Table~\ref{tab:th_selection}). Using those TH lines to derive wavelength solutions, we demonstrate a peak-to-peak precision smaller than 0.75\ms\,for the 13 years of HARPS-N operation.

\subsection{Master flat-fielding}  \label{sect:master_flat_field}

For the previous HARPS-N solar measurements release \citep[][]{Dumusque:2021aa}, the data were reduced using master flat fields.
\edit{These master flat fields are} built by accumulating flat-field frames over several days to avoid being S/N limited by flat-field calibrations when working with daily-binned solar spectra \citep[see Sect. 2.4 in][]{Dumusque:2021aa}. Indeed, as ten flat-field frames are taken daily as part of the calibration sequence, the S/N of those combined frames ($\sim$950 as ten frames \edit{each} with S/N$\sim$300 are used) will be the maximum S/N that can be reached when combining daily solar observations all reduced with the same flat-field calibration. In addition, for this previous data release, rather than using the standard flat-field frames, where both science and calibration fibres are illuminated with a tungsten lamp, we used localisation frames that are composed of two frames with the same tungsten lamp illuminating one fibre at once. Illuminating each fibre separately prevents spectral ghost contamination from the other fibre. Although better in principle, \edit{in practice} the localisation frames are obtained at the beginning of the calibration sequence just after the tungsten lamp is switched on. During this time, the flux of the tungsten lamp varies significantly, which \edit{does not} impact spectral order localisation, but strongly influences flat-field measurements. As a result, the obtained master flat fields were more sensitive to flux variation at the level of the tungsten lamp, rather than pixel-to-pixel response. We therefore decided to remove the use of master flat fields using localisation frames, and reduced the solar data presented in this paper with only the ten available flat-field frames obtained each day. \boldfont{Interested readers about the generation of those master flat fields should have a look at Sect. 2.4 in \citet{Dumusque:2021aa}.}

\subsection{Correcting for the calcium activity index contamination} \label{correcting_calcium}

As for the previous data release, we corrected the derived Mount-Wilson S and $\log{R'_{HK}}$ activity indexes \edit{for} ghost contamination, using the same method as described in Sect.~2.6 in \citet{Dumusque:2021aa}. This method \edit{consists of} extracting locally, on both sides of the orders containing the CaII H and K lines, the flux from ghosts using the same profile as for extracting the orders. Then, an extrapolation of the flux was done to estimate the contaminating flux, which was then subtracted when computing the Mount-Wilson S activity index. \boldfont{Interested readers about this ghost contamination correction for the calcuim activity index should have a look at Sect. 2.6 in \citet{Dumusque:2021aa}.}

\section{Curating and correcting the HARPS-N solar data set from known systematics} \label{RV_data_set}

On July 14th 2015, the HARPS-N solar telescope started routine daily observations of the Sun. Ten years later, as of May 5th 2025, a total of 173793 solar spectra had been recorded. This corresponds to an average of 51 measurements per day, and thus more than 4 hours of daily observations considering the 5-minute exposure time used to average out signals from solar oscillations.

\subsection{Curating the solar data} \label{bad_data}

To reject bad weather data and measurements affected by instrumental instability, we performed the following successive selections. Between January 30th 2024 and July 31st 2024, HARPS-N recorded solar spectra for which the RVs present a jitter of hundreds of \ms due to a misalignment between the solar telescope and its guiding camera. An intervention on May 10th 2024 solved the issue, however, the problem reappeared a few days later. 
\edit{We therefore rejected all data taken between January 30th and July 31st 2024. This represents 4.4 per cent of the full data set (7653 frames).}
We then rejected overall extreme RVs values, by removing data points for which the derived RV is further than 100\ms from the 101.53\ms median values we get for the Sun, in addition to measurements for which the airmass is larger than five. This selection removes an additional 1.6 \% of the measurements (2725 frames).

For the remaining observations, we followed the process presented in \citet{Collier-Cameron:2019aa} and \citet{AlMoulla:2023aa} to reject measurements affected by clouds or strong calima\footnote{\boldfont{Calima is the term used for significant extinction induced by Sahara sand in the atmosphere on top of the Canary islands, where HARPS-N is located}}. The correction is based on modelling the extinction coefficient within each day of measurements, and rejecting measurements that do not follow this \edit{daily model}. Those bad measurements trace lower flux values than expected, induced by clouds, reduced visibility due to calima, or, in very few cases, instrumental problems. We note that HARPS-N stops observing the Sun if the solar telescope guiding camera cannot track it anymore, however, this happens only when the cloud coverage is significant. Thin clouds not evenly distributed over the solar disc will induce a large Rossiter-McLaughlin-type effect, with RVs that can depart from the true value by 
hundreds of meters per second. The Bayesian mixture-model approach presented in \citet{Collier-Cameron:2019aa} returns the probability that the flux measurement follows the linear daily extinction law. We note that such a selection makes the assumption that extinction does not vary significantly over several hours, which can only be a first-order approximation. Nevertheless, we rejected all observations for which the probability was lower than 90\%, which when looking at the stability of RV measurements over several days seemed to be a reasonable threshold. In addition, to remove data that could be contaminated by clouds despite the above selection, we rejected all data for which the photocentre of the measurement was off by 0.5\% from the mid-time of the exposure. This seems a drastic cut, however, it only rejects an additional 2106 measurements. We finally rejected measurements for which the magnitude of the Sun as measured in spectral order 98 (spectrograph order 60, central wavelength 6245\,\AA) was departing by more than 1.4 from the -12.9 median value.
This happened between November 12th 2017 and February 12th 2018 when the fibre transporting light from the solar telescope to the HARPS-N calibration unit was broken, following an instrumental intervention. 
We note that data exist during this time, but with much lower RV precision due to a significant drop in S/N. We rejected those data for consistency. Overall, rejecting observations with lower flux than expected removes 25.0\% of the solar measurements (43494 frames).

\begin{figure}
        \center
	\includegraphics[width=9cm]{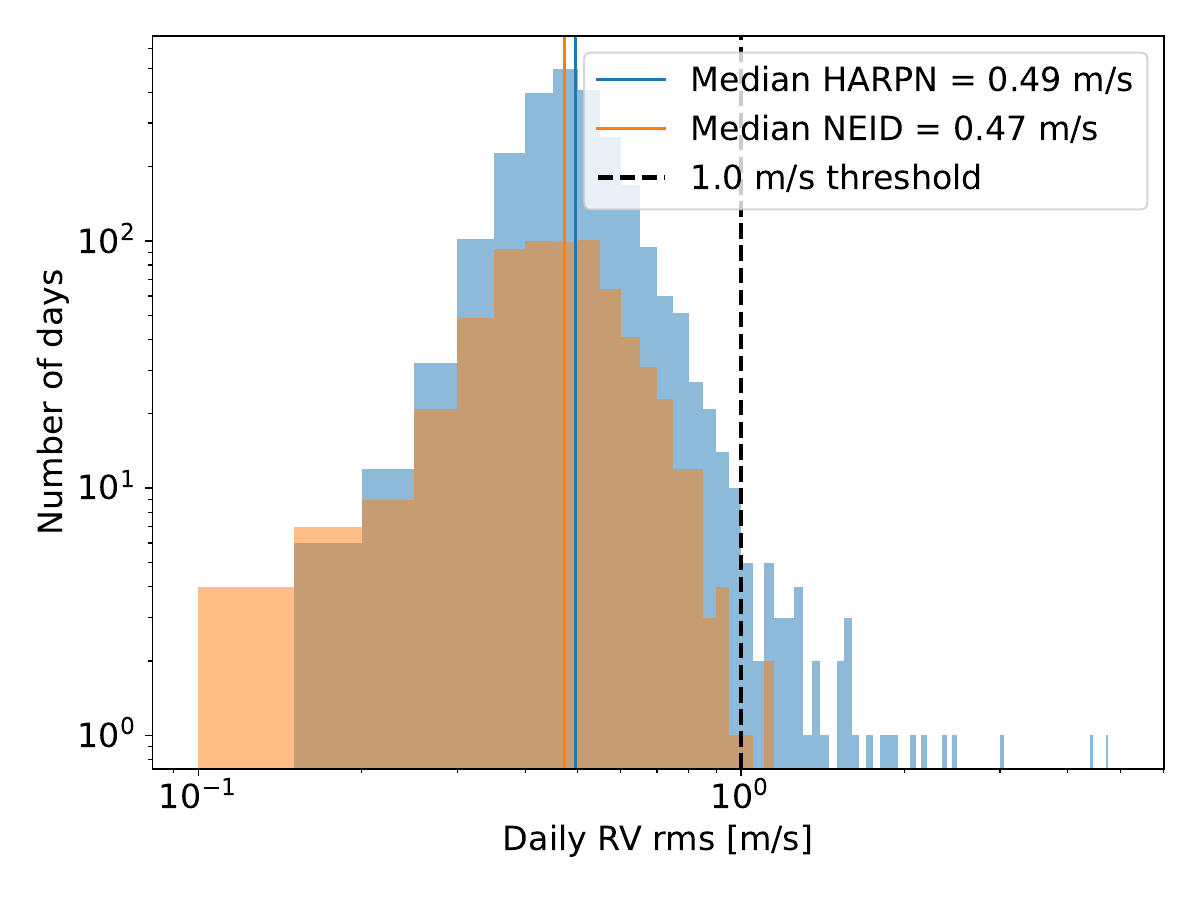}
	\caption{Histogram of the daily RV rms for the HARPS-N solar data presented in this paper, compared with the same statistics from the NEID data published in \citet{Ford:2024aa} for comparison. We note that the NEID data correspond to the 5-minute binned NEID solar data, \emph{rvs\_binned\_5.csv}, available here \url{https://zenodo.org/records/13363762}.} 
	\label{fig:daily_rv_rms}
\end{figure}

After rejecting all those points, we were still left with a few significant outliers of unidentified origins. We therefore fitted the general magnetic cycle trend with a second-order polynomial, \boldfont{measured the median absolute deviation of the residuals, and rejected 88 outliers (0.05\%) that were not compatible with the fitted polynomial within ten times the measured median absolute deviation}. We then finally measured the RV rms for each day of observations with remaining data, and show the corresponding distribution in Fig.~\ref{fig:daily_rv_rms}, alongside the distribution of the same statistics extracted from NEID published solar data \citep[][]{Ford:2024aa}. As a clear drop is observed at 1\ms, we decided to reject all days for which the daily RV rms is above this threshold, which in total rejected 43 days of measurements for a total of 1361 observations.

With this drastic selection we rejected in total 30\% of the obtained solar data. Although this rejection rate might seem high, we prefer to stay on the very cautious side in terms of data curation to deliver to the community the most precise data set as possible. Even with this strict rejection criteria, we still provide to the community a data set of 118472 well-curated data points, over a decade. We note that although the data published with this paper have those cuts applied, it is possible to access all the data and the relevant information to perform specific selections using the DACE API. See Sect.~\ref{data_release} for more information about how to access those data.
%
%

\subsection{Optimised instrumental drift correction throughout the day} \label{drift_corr}

On HARPS-N, as on the majority of high precision RV spectrographs, the zero point of the instrument is recalibrated everyday through the derivation of a new wavelength solution. On this instrument, the wavelength solution is derived from a frame in which the emission spectrum of a TH-AR HC lamp is illuminating the science fibre,  and the emission spectrum of a Fabry-P\'erot \'etalon (FP) is illuminating the calibration fibre \citep[e.g.][]{Wildi-2011}. The Sun will then be observed on the science fibre to assure that the same LSF affects both the solar spectrum and the TH-AR spectrum used for wavelength solution. The calibration fibre during a solar observation will be illuminated with the FP to allow drift measurements of the instrument with respect to the time when the wavelength solution calibration was acquired. Several studies have shown that pressure and temperature controlled Fabry-P\'erot \'etalons can reach a stability well below 0.1\ms per day \citep[see 
e.g. Sect.~5 in][]{Schmidt:2022aa}, therefore suitable for drift measurements over 24 hours. However, this was no longer the case for the HARPS-N FP between January 31st 2018 and May 25th 2019 (JD = 2458150 and 2458629) when, following the replacement of the laser driven light source illuminating the FP cavity, a series of different problems happened on this calibration source. \boldfont{An in-depth analysis is described in Appendix~\ref{app:drift_correction}, and interested readers are invited to look into it.} The main conclusion is that during this time the FP drift could be more than a\ms per day, thus strongly impacting the RV precision of solar observation. This was because before May 27th 2019 (JD = 2458631), no morning calibrations were taken and all data were reduced with wavelength solutions obtained during the preceding afternoon, thus up to 24 hours apart. Although the FP issue was solved after that date and never happened again, we decided to add HARPS-N calibrations every morning before solar data were acquired. 

We also describe in Appendix~\ref{app:drift_correction} a solution to track instrumental drift when the FP was providing erroneous measurements. Considering other ways of measuring instrumental drift, we also describe in that Appendix an optimisation of the daily drift correction by choosing daily between three different drift correction models. To give an idea of the improvements brought by the optimised drift correction, we computed for a few different periods the RV rms of the obtained data and reported the values in Table~\ref{tab:drift_correction}. As an example, we reach a RV rms of 0.84\ms on a 110-day low activity period between BJD 2458640 and 2458750 (5 June 2019 to 23 September 2019).
\begin{table}[h]
  \centering
  \footnotesize
  \begin{tabular}{lc}
    \hline \hline
    \textbf{Drift correction method} & \textbf{RV rms [m/s]} \\ 
    \hline
    Period with unstable FP: BJD 2458500 to 2458700 & \\ 
    \hline
    Simultaneous FP drift correction & 1.13 \\
    Wavelength solution plus high-frequency FP & 1.07 \\\
    Optimised drift correction & 1.03 \\
    \hline
    Low activity period: BJD 2458640 to 2458750 & \\ 
    \hline
    Simultaneous FP drift correction & 0.87 \\
    Wavelength solution plus high-frequency FP & 0.94 \\
    Optimised drift correction & 0.84  \\
    \hline
    Medium activity period: BJD 2457589 to 2457660 & \\ 
    \hline
    Simultaneous FP drift correction & 1.25 \\
    Wavelength solution plus high-frequency FP & 1.24 \\
    Optimised drift correction & 1.17 \\
  \end{tabular}
  \caption{RV rms for different periods of the solar data using different instrumental drift correction methods. We note that the simultaneous drift correction method includes the correction when the FP drift was unstable (see Fig.~\ref{fig:drift_diff}). The optimised drift correction method corresponds to the best one among all the ones tested on the solar data (see text).}
  \label{tab:drift_correction}
\end{table}
%
%
%

\subsection{Blaze variation correction due to cryostat contamination} \label{sect:cryo_trends}

On HARPS-N, a leak at the level of the detector cryostat \edit{has} imposed a periodic warm-up cycle of the device every six months on average, as can be seen in Table~\ref{tab:intervention_details}. This leak \edit{increased} the level of the ghosts on HARPS-N raw frames over time, contaminating, for example, the measurement of the calcium activity index, as demonstrated in \citet{Dumusque:2021aa}. Each warm-up cycle was resetting this contamination to a minimum level. As discussed \boldfont{in detail} in Appendix~\ref{app:blaze_corr} we discovered thanks to the solar data that this contamination was also affecting the derivation of the CCF and with it the RV. The solution described in Appendix~\ref{app:blaze_corr} to correct for this systematic will be tested further in the next section.

\subsection{Optimisation of the correction for TH lamp drift and the blaze variation corrections} \label{sect:opti_th_drift}

\begin{figure}
        \center
	\includegraphics[width=9cm]{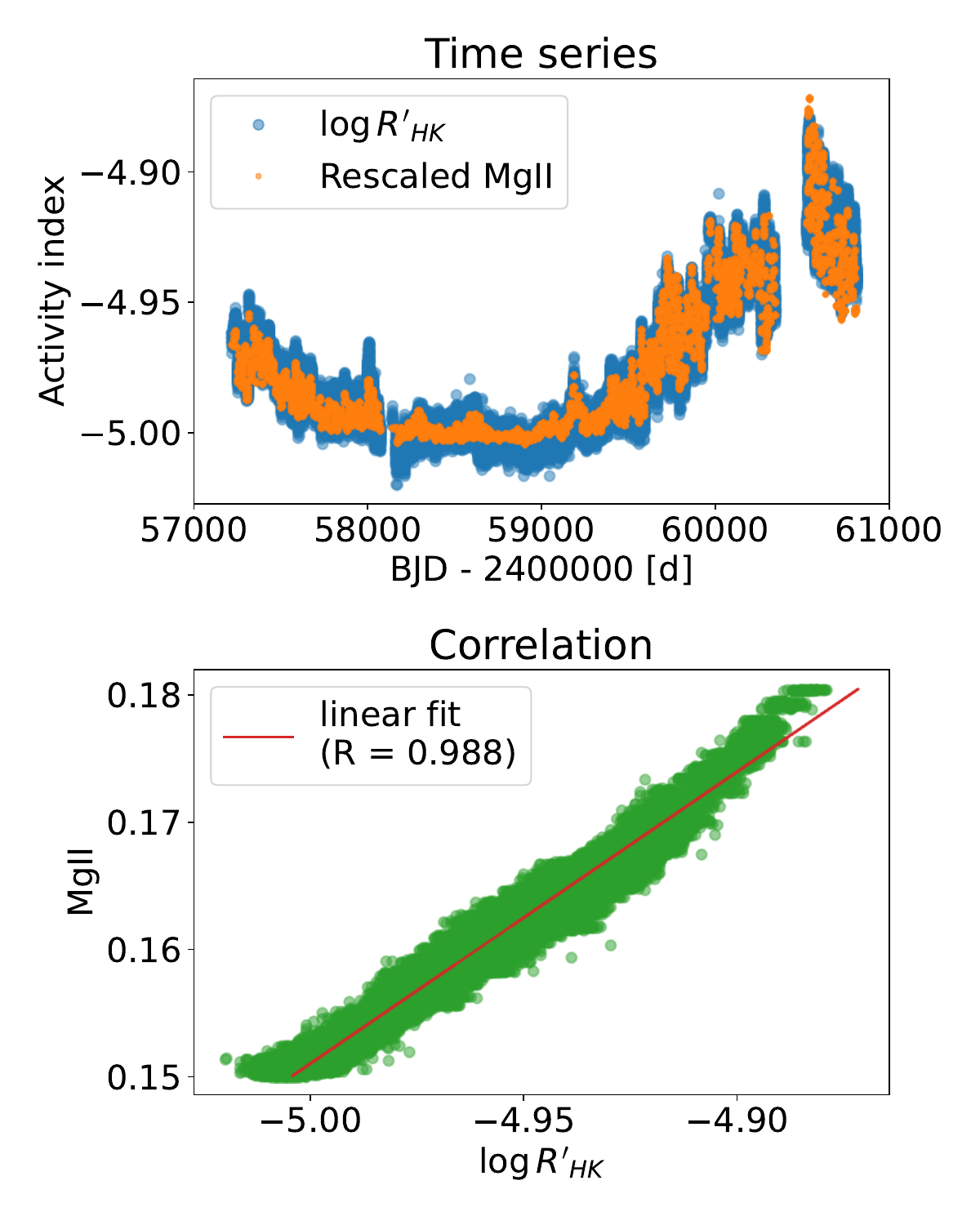}
	\caption{\boldfont{\emph{Top:} Comparison between the $\log{R'_{HK}}$ activity index derived from HARPS-N and the linearly rescaled Bremen Composite Mg2 activity index. \emph{Bottom:} Correlation between those two activity indices, that demonstrate their very strong relation.}} 
	\label{fig:mg2_rhk}
\end{figure}

In Sect.~\ref{sect:th_drift}, we discussed our strategy to correct for the drift of thorium lines induced by lamp aging. By comparing the ratio of the AR line offset to TH line offset at several lamp changes, we conclude that AR1 lines shifts on average 20.95 times more than TH lines. However, this measurement is done each time between an old dying lamp and a new one and applying this correction at all times by measuring AR1 line drift assumes that the lamp ages linearly through time, which is probably not correct. Although the solar RVs are not stable on the long-term due to the magnetic cycle, those RVs should be strongly correlated on long-timescales to the Mount-Wilson S-index \citep[][]{Wilson-1968} and related $\log{R'_{HK}}$ \citep[][]{Noyes-1984} activity indices \citep[e.g.][]{Dumusque-2011c, Meunier:2020aa}. Imposing such a correlation can be used to test how well our TH lamp aging correction performs. \boldfont{HARPS-N observes down to 3875 \AA\ and therefore gets access to the Ca II H \& K lines at  3969 and 3934 \AA\ used to derive $\log{R'_{HK}}$. However, we wanted to use an activity index independent from HARPS-N spectra, to prevent any long-term instrumental systematics that could affect both RVs and activity index to be hidden when correcting RVs from this activity index. To the best of our knowledge, daily monitoring of the solar calcium activity index stopped a decade ago, with the end of the US National Solar Observatory Evans Solar Facility in 2015 and shutdown of SOLIS in 2017 for a planned relocation to the Big Bear Observatory. Therefore, we selected the Bremen Composite Magnesium II (Mg2) activity index}\footnote{\url{https://www.iup.uni-bremen.de/UVSAT/data/}}\boldfont{, which, similarly to $\log{R'_{HK}}$, is a measure of the flux emitted in the cores of the Mg2 H \& K components at 2796 and 2803 \AA. The cores of those lines are formed only slightly higher up in the chromosphere than the Ca II H \& K line cores, which explains the strong correlation observed between the Mg2 activity index and $\log{R'_{HK}}$ shown in Fig.~\ref{fig:mg2_rhk}.}

\begin{figure*}
        \center
	\includegraphics[width=18cm]{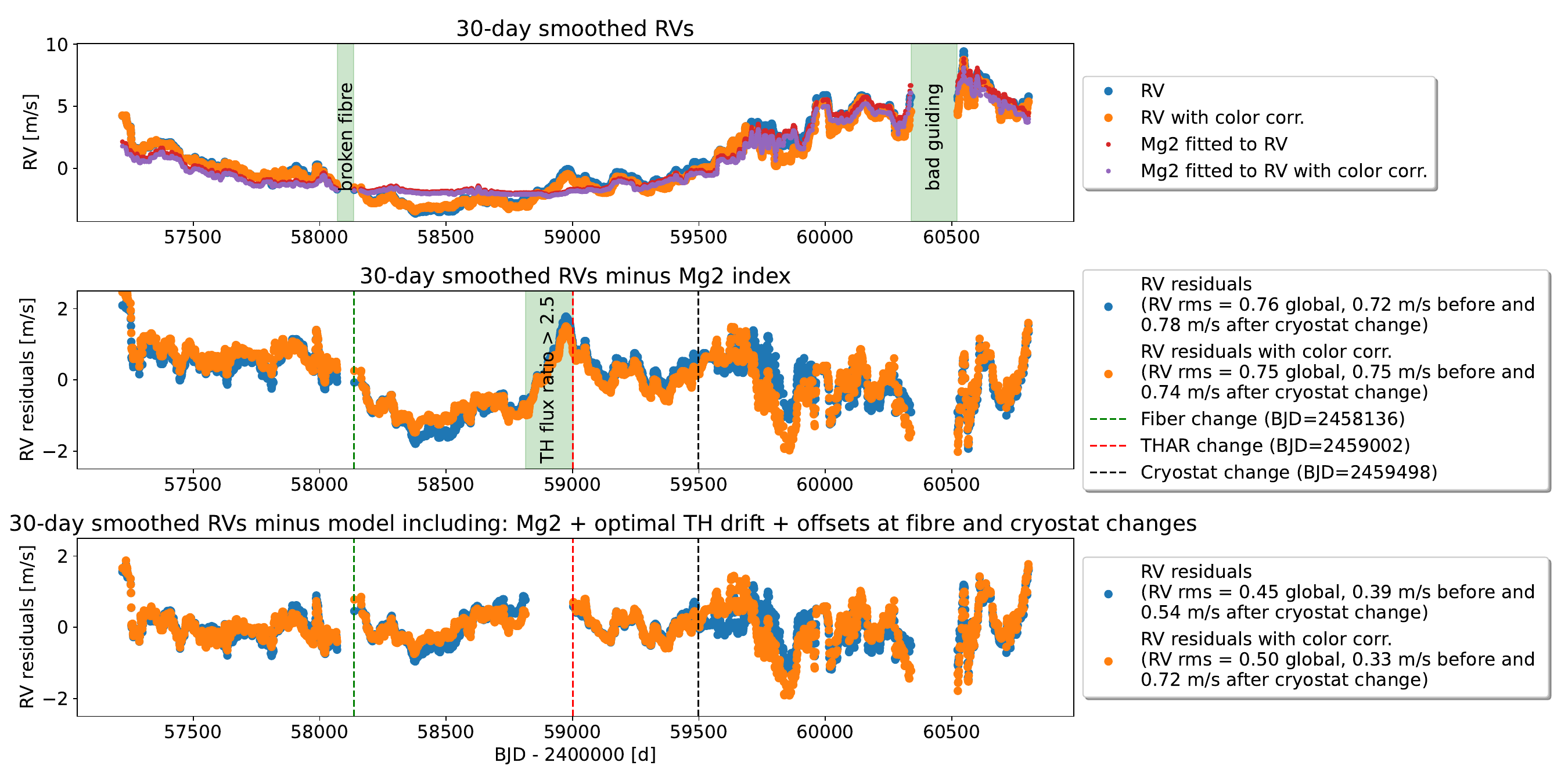}
	\caption{\emph{Top:} RVs without and with the correction for blaze variation, smoothed using a 30-day moving average to mitigate stellar activity signals due to the Sun's rotation (in blue and orange, respectively). The red and purple dots corresponds to the best-fitted linear model of the Mg2 activity index to those two different time-series. \emph{Middle:} RV residuals after subtracting the best Mg2 activity index linear model. The vertical lines highlight the change of the broken solar feed fibre, the change of the TH-AR HC lamp used for wavelength solution and the change of the detector cryostat (see legend). The green region highlights a period when the flux of the TH-AR HC lamp was larger than 2.5 times it nominal value (see Fig.~\ref{fig:th_ar_drift}). \emph{Bottom:} RV residuals after subtracting a linear model including the Mg2 activity index, two offsets for the broken fibre and the cryostat change and a scaling factor for the AR1 to TH ratio.} 
	\label{fig:color_and_mg2_corr}
\end{figure*}

To test how well our TH lamp aging correction performs, we therefore linearly fitted the Mg2 activity index to the HARPS-N RVs, after both time series were smoothed using a 30-day moving average\footnote{algorithm used: pandas.rolling("30d", min\_periods=1, center=True).mean().} to mitigate the effect of solar rotation and focus on the long-term trend. In the top panel of Fig.~\ref{fig:color_and_mg2_corr}, we can see the solar RVs for which the wavelength solutions were corrected for using an AR1 to TH ratio of 19.95. When correcting those RVs with a linear fit to the Mg2 activity index, the residuals shown in the middle panel of the same figure present \boldfont{significant systematics with a RV rms of 0.75 \ms. Looking to the systematics in more details, we see a behavior similar to the AR line drift as a function of time (see middle panel of Fig.~\ref{fig:th_ar_drift}). This tells us that the correction for the thorium drift by using the behavior of AR1 lines is not optimal (see Sect.~\ref{sect:th_drift}) and therefore,} we decided to include the AR1 to TH ratio as a free parameter \boldfont{and fit the value of this ratio using the solar data themselves.} In addition we also decided to include in the fit of the RVs two offsets for the change of the solar feed fibre at BJD = 2458138 (19 January 2018) and the change of the HARPS-N CCD cryostat at BJD = 2459498 (10 October 2021). \boldfont{These two interventions changed the PSF of the spectrograph on the detector, once by changing the entrance illumination of the spectrograph, the other by changing the focus of the image on the detector as it was moved. We therefore expect an offset of the RVs during these interventions.} We also rejected all data between BJD = 2458815 (27 November 2019) and the TH lamp change on BJD = 2459002 (1 June 2020), as during this time the TH-AR HC lamp used for wavelength solution was dying and its overall flux (what we call the TH flux ratio) was higher than 2.5 times its nominal value, as seen in Fig.~\ref{fig:th_ar_drift}. We see that this deviation from nominal value induces a large drift at the level of the RV residuals shown in the middle panel of Fig.~\ref{fig:color_and_mg2_corr}. This RV drift are likely due to significant changes at the level of the TH spectrum due to a much larger flux, which affects wavelength solution derivation, but does not track the drift of the spectrograph. A possible explanation is that with larger flux, the blending of the TH lines used for wavelength solution derivation will change, which modifies lines asymmetry and in the end induces a spurious RV drift. The obtained RV residuals after removing this more complex model, as seen in the bottom panel of Fig.~\ref{fig:color_and_mg2_corr}, are now much flatter on the long term. We note that we performed this analysis for both the original RVs extracted from the ESPRESSO DRS and the RVs corrected from the blaze variation induced by cryostat contamination (see preceding section). After correcting for the full model including the Mg2 index, the AR1 to TH ratio and the two offsets, the RV rms of the residuals before the change of the cryostat on BJD = 2459498 (10 October 2021) is 0.39 and 0.33\ms, and 0.54 and 0.72\ms after, respectively. Thus, the blaze variation correction gives better results only for the data before the cryostat change. For \edit{data after this change}, the larger RV rms is mainly induced by an unexplained systematic observed between approximately BJD = 2459700 and 2459900 (April 30th to November 16th 2022). Given the significantly large RV rms for the blaze-corrected RVs after the cryostat change, we decided not to apply the correction for those respective data. In the end, the best fit converges to an offset of 1.21\ms when the solar feed fibre was changed, an offset of 0.86\ms when the cryostat was changed, an RV rms of 0.42\ms and an AR1 to TH ratio of 15.83. This AR1 to TH ratio value is different from the 19.95 one found by comparing old and new lamps and therefore calls into question the linearity between AR1 and TH lines drift as a function of time. This discrepancy was already discussed in more detail at the end of Sect.~\ref{sect:th_drift}.
%
%

\boldfont{Although the choice of model with an optimisation of the AR1 to TH ratio and two offsets corresponding to significant interventions is well justified, we discuss here the impact of each adjusted parameter. The magnetic cycle can first be adjusted using a linear correlation with the Bremen Composite activity index or the \logrhk derived by HARPS-N. The RV rms obtained from the residuals is 0.75 and 0.71 \ms , respectively. Adding the AR1 to TH ratio optimisation lowers the RV rms to 0.60 \ms\ in both cases. Adding on top the two offsets lowers the RV rms even further to the level of 0.45 \ms.}

%


\section{Results} \label{results}

\subsection{The curated solar data} \label{curated_RVs}

\begin{figure*}
        \center
	\includegraphics[width=18cm]{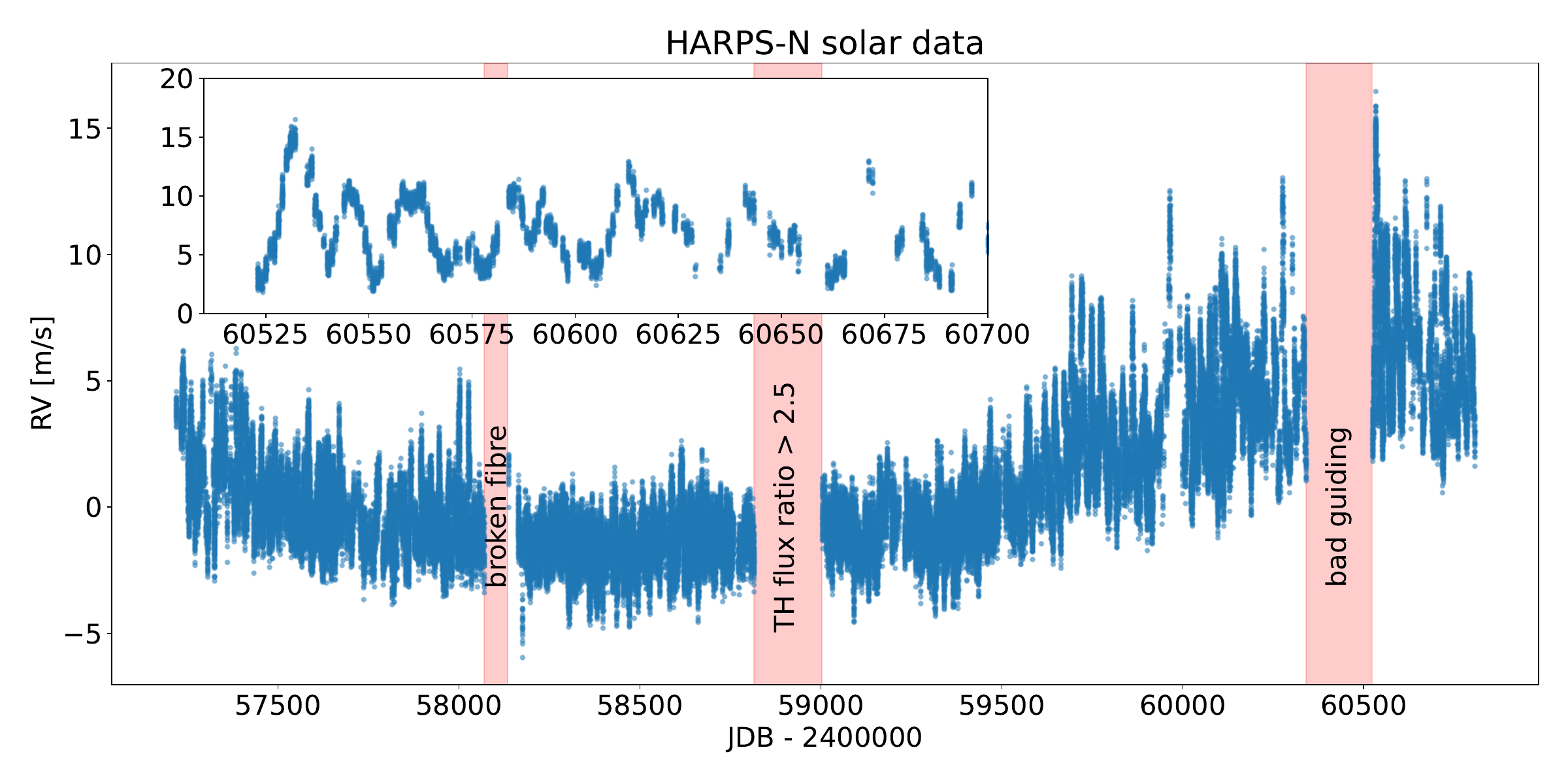}
	\caption{RVs of the carefully curated HARPS-N solar data as a function of time. The three important gaps highlighted by red regions correspond to a broken fibre guiding light from the solar telescope to the HARPS-N calibration unit, a moment in time where the TH-AR HC lamp was out of specification to provide extremely precise wavelength solutions and a problem of alignment between the guiding camera tracking the Sun and the solar telescope itself. The inset focuses on nearly 200 days of solar RVs just after solar maximum, where rotational modulation, induced by active regions co-moving with the solar surface, is very strong.}
	\label{fig:solar_RVs}
\end{figure*}

We show in Fig.~\ref{fig:solar_RVs} the curated HARPS-N solar RVs, which contain a total of 109466 good quality observations out of 173793 in total, from July 14th 2015 to May 5th 2025. We see three important gaps in those data, around BJD = 2458100, 2459900 and 2460450.
\edit{These gaps are due to, in order, }a broken fibre between the solar telescope and the HARPS-N calibration unit, the flux of the thorium lamp going significantly above its nominal value and misalignment between the solar telescope and the guiding camera used to track the Sun. The RV rms over this decade of observation is 2.95\ms.
\edit{This variability is} dominated by the long-term variation induced by the solar magnetic cycle. The inset in this top panel highlight nearly 200 days of observation just after the solar magnetic cycle maximum, where the RV effect induced by magnetic region rotating with the Sun is clearly visible. The median photon-noise RV precision for those observations is 0.28\ms. The median intraday RV rms is 0.49\ms as can bee seen in Fig.~\ref{fig:daily_rv_rms}.

\begin{figure*}
    \includegraphics[width=18cm]{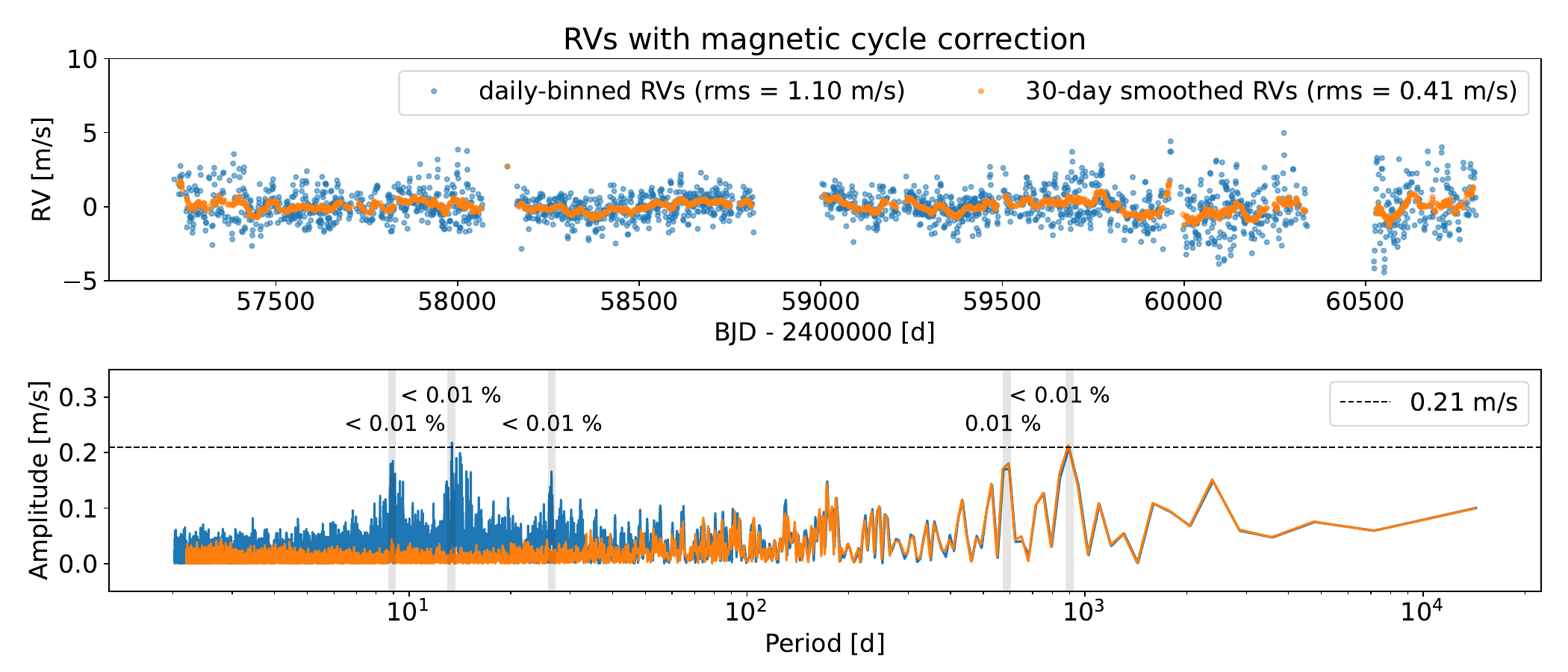}
    \caption{\emph{Top:} Daily-binned RVs and 30-day smoothed average once a linear fit to the Mg2 activity index has been removed to mitigate the RV long-term effect induced by the solar magnetic cycle. \emph{Bottom:} Corresponding Lomb-Scargle periodograms in RV amplitude. As we can see, the 30-day smoothing absorbs signal up to 150 days and both data sets give the same signals above 150 days. The horizontal dashed line corresponds to the maximum peak for periods longer than 150 days, which is seen at a period 895 days, with an RV amplitude of 0.21\ms. The numbers on top of each significant peak, highlighted in gray, correspond to the signal false alarm probabilities.}
	\label{fig:solar_RVs_without_magn_cycle}
\end{figure*}

As we can see in the top panel of Fig.~\ref{fig:solar_RVs_without_magn_cycle}, once the long-term trend induced by the solar magnetic cycle is removed using a linear fit to the Mg2 activity index, the daily-binned RVs show an RV rms of 1.1\ms. This value decreases to 0.41\ms once rotational modulation has been smoothed out using a 30-day moving average\footnote{algorithm used: pandas.rolling("30d", min\_periods=1, center=True).mean().}. \boldfont{We note that repeating this exercise by using the \logrhk derived from HARPS-N instead of the Mg2 activity index gives very similar RV rms with 1.12 and 0.44 \ms , respectively. Once the effect from the magnetic cycle is modeled,} the long-term RV precision of the HARPS-N solar data is therefore below the 0.5\ms level. At this extreme precision, we see an unknown systematic between approximately BJD = 2459800 and 2460000 (August 8th 2022 and February 24th 2023). Although understanding this behavior is out of the scope of the present paper, it should be considered in further analysis. 

In the bottom panel of Fig.~\ref{fig:solar_RVs_without_magn_cycle}, we show the Lomb-Scargle periodograms of the daily-binned and 30-day smoothed RV time series. Rather than showing the periodogram normalised power as a function of period, we show in that panel the periodogram amplitude in \ms, to give an idea of planetary signal detectability as a function of period. As we can see, the smoothing mitigates signals up to 150 days, but for longer periods both data sets give very similar results. For those longer periods, we see two significant peaks (false alarm probability of 0.01 and smaller than 0.01) at 585 and 895 days with an amplitude of 0.18 and 0.21\ms, respectively. Coincidentally, the signal of Venus that we would see in the HARPS-N solar data would appear at a period of 584 days, the synodic period of Venus, matching perfectly with one of these signals. However, all the data presented in this paper are derived in the heliocentric rest-frame and thus do not contain any signal from Solar System planets. On a positive note, we reach a planetary detection limit of 0.21\ms for all periods above 150 days, including the habitable zone of solar-type stars, without any post-processing of the HARPS-N solar data. This limit would correspond to a 2.5 Earth-mass planet with an orbital period of one year around the Sun.

%
%
%
%

\subsection{Comparison with NEID solar data} \label{compa_NEID}

\begin{figure*}
    \center
    \includegraphics[width=18cm]{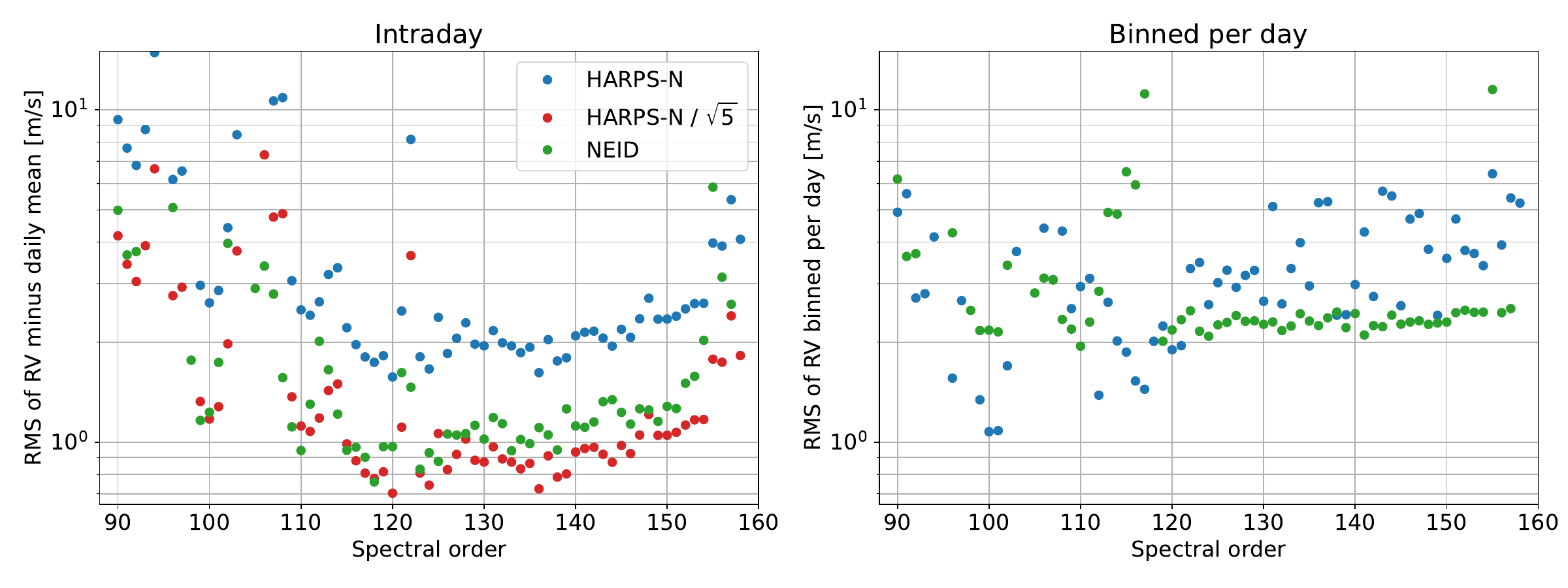}
    \caption{Intraday RV rms on the left and RV rms of data binned per day on the right for each spectral order. Blue dots corresponds to an analysis of the HARPS-N solar data presented in this paper, and green dots to the analysis of the NEID solar data shown in Fig.~5 in \citet{Ford:2024aa} (data courtesy of Eric Ford). The left panel focuses on short-term precision, and is obtained by removing from each day of observation its daily mean, and the right panel focuses on long-term precision, dominated by the long-term systematics induced by the solar magnetic cycle. \boldfont{For the intraday analysis (left panel), we should compare the green NEID data with the HARPN red dots labeled \emph{HARPN / $\sqrt{5}$} that corresponds to the blue dots divided by the square root of five. This allows to compensate for the S/N difference between HARPS-N (S/N$\sim$300 over 5 minutes) and NEID (5 observations of S/N$\sim$300 over 5 minutes) solar data}.}
	\label{fig:solar_RVs_intra_and_daily_binned}
\end{figure*}

In Fig.~\ref{fig:solar_RVs_intra_and_daily_binned} we show the intraday RV rms and the rms of daily-binned RVs for each spectral order of HARPS-N. To obtain the intraday statistics, we removed from each day of observation its mean RV value, making sure to only take days for which at least 15 measurements were obtained. Those results can be compared with the same ones obtained from NEID solar data \citep[see Fig.~5 in][]{Ford:2024aa}, \boldfont{that we reported as green dots in our figure}. We note that the intraday RV rms is around 2\ms for most spectral orders, except for the redder ones (spectral orders 110 to 90) where the RV content is lower due to fewer spectral lines and also more telluric contamination. \boldfont{In comparison, the NEID results show an intraday RV rms per order, which is on average a factor of two lower. This discrepancy can be explained by the difference in S/N of the HARPS-N and NEID data. In five minutes, HARPS-N obtains one measurement\footnote{We note that a neutral density filter is used to attenuate the light entering the spectrograph to prevent detector saturation during the 5-minute exposure time used to mitigate stellar oscillations. Without such a filter, saturation would be reached after 20 seconds.} of the Sun with a S/N of $\sim$300. NEID obtains such a S/N in a 55 second exposure, but the data analyzed in \cite{Ford:2024aa} are obtain after binning five consecutive measurements, thus corresponding to 4.6 minutes on the Sun for a wall clock time of 7.12 or 6.45 minutes (before and after September 16, 2021). Given this binning, stellar oscillations are equally mitigated in HARPS-N and NEID data, at a level of a dozen \cms. In addition, the other stellar signals over one day, granulation and supergranulation should impact HARPS-N and NEID data in a similar way and would imply a maximum RV rms at the level of 0.7 \ms\ \citep[e.g.][]{AlMoulla:2023aa}, which is lower than the minimum intraday RV rms of 0.8 \ms obtained by NEID for physical order 118. Therefore, the intraday RV rms per order for NEID and HARPS-N are photon-noise limited. To compare the intraday precision of HARPS-N and NEID, we therefore have to compare data at similar S/N, which can be obtained by dividing the intraday RV rms of HARPS-N by the square root of five. When doing so,} the HARPS-N and NEID intraday RVs rms per order are \boldfont{at a similar level}. This is also the case when averaging the RVs over all orders as the median intraday RV rms for the HARPS-N solar data is 0.49 \ms, to be compared with 0.47 \ms for NEID (see Fig.~\ref{fig:daily_rv_rms}). HARPS-N and NEID thus demonstrate a very similar level of stability over one day.

Looking now at the rms of the daily-binned RVs on the right panel of Fig.~\ref{fig:solar_RVs_intra_and_daily_binned}, we are no longer dominated by photon noise, as the median number of observations taken each day is 53. In this case, the RV rms is between 2 to 5\ms for the majority of orders. In comparison, for NEID the majority of orders gives RV rms between 2 and 3\ms. First, NEID solar observations presented in \citet{Ford:2024aa} only cover 3 years of observations, without going through solar maximum. In our case, we have 10 years of data covering the recent solar maxima, in addition to a part of the previous solar cycle. The larger daily-binned RV rms per order for HARPS-N solar data can therefore be explained by this longer baseline of the observation. This is seen as well on the global RVs, where HARPS-N data give an RV rms value of 2.95\ms while it is only 2.11\ms for NEID data. However, for the bluest orders (spectral orders 160 to 150) we reach a RV rms of up to 5-6\ms while NEID stays below 3\ms. This discrepancy on the blue side could be linked to the observed variation of spectral ghosts in HARPS-N, that are linked to a leak at the level of the detector cryostat before its replacement on BJD = 2459498 (10 October 2021). In Sect.~\ref{sect:cryo_trends} we provide a correction for the change in blaze induced by these variations, but we do not correct the spectra for the varying contaminating flux of ghosts. The larger RV rms observed for the red spectral orders are likely due to telluric contamination, as also seen in NEID.

\begin{figure*}
        \center
 	\includegraphics[width=18cm]{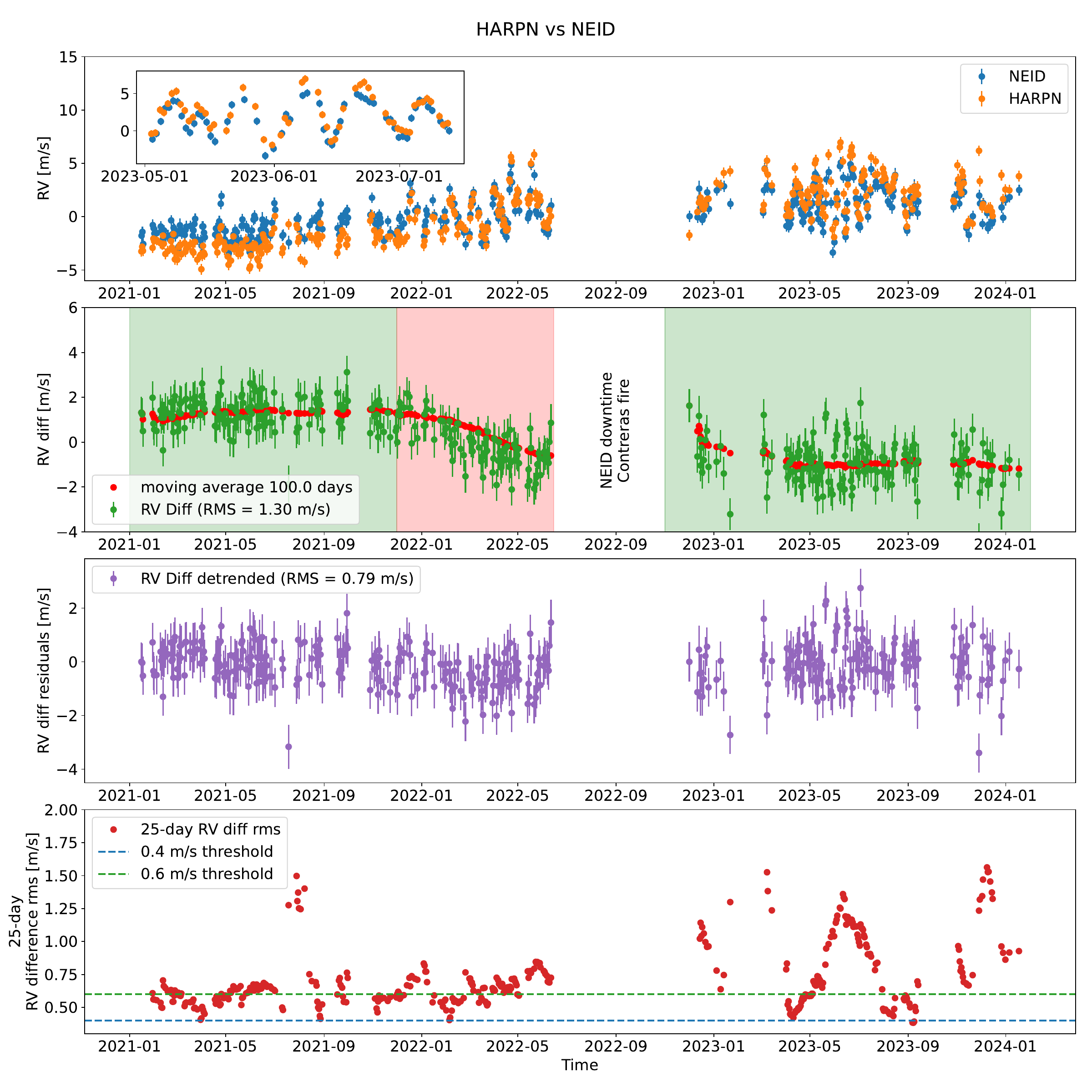}
	\caption{\emph{1st panel:} Comparison between the HARPS-N and NEID solar RVs published in \citet{Ford:2024aa}. The inset zooms on two months of the data where rotational modulation from active regions is clearly visible. \emph{2nd panel:} RV difference between those two data sets that presents an RV rms of 1.3 \ms. The red curve corresponds to a moving average with a period of 100 days. \emph{3rd panel:} RV difference detrended by the moving average that presents an RV rms of 0.79 \ms. \emph{4th panel:} Moving standard deviation of the RV difference residuals with a window of 25 days to be sensitive to solar rotation.} 
	\label{fig:HARPN_NEID_comparison}
\end{figure*}

To test further the solar data obtained by HARPS-N and NEID, we compare the daily-binned HARPS-N and NEID RV time series in the first panel Fig.~\ref{fig:HARPN_NEID_comparison}. We only show in this figure data that have been taken during the same days. Although HARPS-N and NEID data are not contemporaneous due to the different longitudes of the instruments \citep[e.g.][]{Zhao:2023ab}, we do not expect RVs to be more discrepant than 1\ms as the main signal on timescales shorter than a day is supergranulation, which is estimated to be of the order of 0.7 to 0.9\ms \citep[e.g.][]{Meunier-2015,AlMoulla:2023aa,Lakeland:2024aa}. Supergranulation has a time scale of up to two days, however, \cite{Lakeland:2024aa} demonstrate (see their Fig.~7) that already on the 8-hour difference between the HARPS-N and NEID data, supergranulation is expected to induce an RV rms of $\sim 0.6$\ms. The second panel of Fig.~\ref{fig:HARPN_NEID_comparison} shows the difference in RVs between the two instruments with an RV rms of 1.3\ms. However, this RV difference, shows an interesting pattern, with periods highlighted in green where the RV difference between the data from the two instruments is flat, and a period (shown in red) before the NEID downtime due to Contreras fire where a significant slope is observed. Removing the long-term trend with a 100-day moving average\footnote{algorithm used: pandas.rolling("100d", min\_periods=1, center=True).mean().} to preserve rotational modulation gives us RV residuals that present RV rms of 0.79\ms, as seen in the third panel of Fig.~\ref{fig:HARPN_NEID_comparison}. In the fourth panel, to better highlight how the jitter of the RV difference residuals varies as a function of time, we computed a moving standard deviation with a window of 25 days to be sensitive to solar rotation. During the first year of observation, where solar activity was close to minimum and therefore the RV jitter should be dominated by supergranulation, we observe an average RV rms value of 0.6\ms, fully compatible with what is expected from supergranulation considering the 8-hour time difference between the two datasets. For the last year, when the Sun was really active, the RV rms can go up to 1.5\ms, which is likely explained by stellar activity.

The observed trend in the RV difference of the HARPS-N and NEID solar data is puzzling, as it significantly calls into question the stability of the spectrographs on the long-term and therefore puts a barrier to the detection of tiny planetary signals on long-period orbits, such as other Earths. However, looking at the raw RV time series (first panel of Fig.~\ref{fig:HARPN_NEID_comparison}), it seems that although solar activity is increasing during the entire time span, the RV difference between the two instruments during the green periods seems flat, while we observe a non-negligible slope during the red period. Also, NEID RVs are higher than HARPS-N RVs at the beginning of the sequence and lower towards the end. It could therefore be that NEID RVs are less sensitive to the effect induced by the solar magnetic cycle. The analysis of the 25-day moving standard deviation of the RV difference between the two datasets in the fourth panel of Fig.~\ref{fig:HARPN_NEID_comparison} also seems to demonstrate a different sensitivity to magnetic activity, as the RV rms difference reaches very high values during high activity. Although the NEID reduction uses the same cross-correlation mask as HARPS-N solar RVs (the ESPRESSO G2 CCF mask), the weighting between orders performed in \citet{Ford:2024aa} is different and could lead to a difference in RV sensitivity to the solar magnetic cycle \cite[see for example Figure 10 in][]{AlMoulla:2022aa}. A more detailed analysis is out of the scope of the present paper, but needs to be investigated further, mainly when more carefully curated solar data from NEID will become available.
%

\subsection{Data release} \label{data_release}

The curated solar data presented in this paper are released to the community thanks to the DACE web interface with DOI indentifier 10.82180/dace-h4s8lp7c. \boldfont{The table with the different time series, including the RVs, activity indices and observational information is also available at ViziR}. More information on those data can be found in Appendix~\ref{app:data_release}.

\section{Conclusion} \label{Conclusion}

In this paper, we describe the reduction and curation of a decade of HARPS-N solar observations. The corresponding high-fidelity spectra and extreme-precision RVs are made publicly available via the DACE web interface (\url{https://dace.unige.ch/openData/?record=10.82180/dace-h4s8lp7c}). See Appendix~\ref{app:data_release} for more information about the released products and please make reference to DOI indentifier 10.82180/dace-h4s8lp7c and this paper when using those data.

The HARPS-N solar data presented in this paper have been reduced using version 3.2.5 of the ESPRESSO DRS. Compared to version 2.2.3, which was used to reduce the data released in \citet{Dumusque:2021aa}, this newer version includes a correction for the drift of TH lines over time due to HC lamp aging. By analysing the behavior of TH and AR lines at each TH-AR lamp change in HARPS-N and HARPS calibrations (see Appendix~\ref{app:th_ageing}), we conclude that TH and AR lines drift in velocity as the lamp ages. We then discuss in Sections~\ref{sect:th_drift} and \ref{sect:opti_th_drift} a correction for the drift of TH lines, which provides long-term wavelength solution stability and thus RV precision. Version 3.2.5 of the ESPRESSO DRS also includes for HARPS-N a new curation of TH lines used for wavelength solution derivation. The final line list used includes a total of 2020 TH spectral lines well distributed over the HARPS-N detector (see Table~\ref{tab:th_selection}), which allows us to perform a robust 2D polynomial fit for wavelength solution derivation \citep[][]{Dumusque:2021aa}. The combination of the TH drift correction and the new selection of TH lines significantly improves the long-term stability of the HARPS-N wavelength solutions, with a measured peak-to-peak deviation smaller than 0.75\ms\,over the 13 years of HARPS-N calibrations.

After reducing the HARPS-N solar data with version 3.2.5 of the ESPRESSO DRS, a total of 173,793 spectra are available from July 14th 2015 to May 5th 2025. A significant fraction of those are, however, affected by the passage of clouds, calima, and known instrumental systematics. We discuss in Sect.~\ref{bad_data} the curation performed to remove spectra with bad RV measurements, which in total rejects 30\% of the data set. By analysing those curated data, we were able to probe a significant systematic induced by instabilities of the FP calibrator used for instrumental drift measurement. As discussed in Sect.~\ref{drift_corr}, we correct for this effect by measuring the drift as a linear interpolation between daily wavelength solutions plus high-frequency FP variations with periods smaller than 4 hours. We also analysed the day-to-day RV variations for the entire solar data set, and for each selected the best drift-correction method between i) FP-based, ii) linear interpolation between wavelength solution plus high-frequency FP variations or iii) no drift correction. As summarised in Table ~\ref{tab:drift_correction}, this optimisation of the drift correction allows us to reach better RV precision. We also discovered that the significant variation of spectral blaze through time, induced by a leak at the level of the HARPS-N detector cryostat, was affecting the RVs derived by the ESPRESSO DRS. This blaze variation changes slightly the flux balance across each spectral order, therefore, changing the weight applied to each spectral line when building the CCF and thus influencing the derived RVs. We discuss a correction for this effect in Sect.~\ref{sect:cryo_trends} and encourage any team working with EPRV spectrographs to include such a correction in their DRS if not already considered.

After rejecting additional data affected by wavelength solutions that cannot be trusted because the flux of the TH-AR HC lamp used to derive them was too high (see Sect.~\ref{sect:opti_th_drift}), and correcting for the systematics discussed in the preceding paragraph, we obtain a total of 109,466 high-fidelity spectra and extremely precise RVs over a decade. The median photon-noise RV precision of those observations is 0.28\ms and the median intraday RV rms is 0.49\ms. On the full data set, the RVs rms is 2.95\ms, a value dominated by the long-term variation induced by the solar magnetic cycle. Once we correct for the magnetic cycle effect using a linear correlation to the Bremen Composite Magnesium II activity index \boldfont{(or HARPS-N \logrhk\ ), we reach an RV rms of 1.1 (1.12)\ms, which is now dominated by the rotational modulation of active regions. After averaging-out the effect of solar rotation using a 30-day moving average, the RV rms decreases to 0.41 (0.44)\ms}. This value demonstrates the extreme stability of the HARPS-N solar RVs over a decade \boldfont{once the magnetic cycle is modeled}. When analysing the periodogram of the 30-day smoothed data, the two most significant signals left in the data are at 585 and 895 days, with amplitudes of 0.18 and 0.21\ms, respectively. These solar data therefore demonstrate that with HARPS-N, we can be sensitive to planetary signals as small as 2.5 times the Earth-mass for periods up to the habitable zone of solar-type stars. Although not yet the 0.09\ms required to detect another Earth, those data do not include any sophisticated post-processing technique to mitigate remaining stellar and instrumental systematics. \boldfont{We note that contrary to what we could think, HARPS-N daily-binned solar data are not at unrealistic S/N compared to what can be obtained for other stars, as a strong limitation is coming from the spectrograph calibrations. The S/N per pixel is limited by the ten frames at S/N $\sim$300 taken for flat-fielding. Therefore, the S/N of daily-binned solar data are at the level of $\sim$950 ($300\,\times \sqrt{10}$). Stellar observations are often carried out with an integration time (including sub-exposure if too bright) of 15 minutes to average stellar oscillations and optimize telescope overheads. We can thus obtain stellar data with similar precision if within 15 minutes, we can reach S/N=950. Using HARPS and ESPRESSO ESO exposure time calculator, considering standard observing conditions (0.8 arcsec seeing and airmass 1.2), this S/N can be obtained in 15 minutes on a V=4.8 and 6 magnitude star, respectively. This limits the number of objects that can be observed at a precision similar to what is obtained for the Sun with HARPS-N, however, ESPRESSO could still observe at such precision 85\% (140) of the stars selected as best candidate for the future Habitable World Observatory mission (Mamajek \& Stapelfeldt 2023\footnote{\url{https://exoplanetarchive.ipac.caltech.edu/docs/2645\_NASA\_ExEP\_Target\_List\_HWO\_Documentation\_2023.pdf}}). Following the HARPS-N solar results and the fact that similar precision can be reached for other stars, we believe} that Earth-twin detections are within reach of the RV technique.

Finally, we compare 3-years of contemporaneous HARPS-N and NEID solar data \citep[][]{Ford:2024aa}. We find very good agreement between the two data sets in terms of short-term precision, however, we find an unexpected linear trend in the RV difference over a 6-month interval. This systematic effect results in a large RV rms of 1.3\ms in the daily-binned RV difference between the two instruments. Once this trend is corrected for, the RV rms over 3 years decreases to 0.79\ms. A more detailed analysis of the RV difference rms between the two datasets demonstrate that during low activity phase, HARPS-N and NEID data are compatible within 0.6\ms, which is the noise expected from supergranulation given the 8-hour time difference between the two data sets. Although out of the scope of the present study, understanding the origin of the trend in RV difference is critical, as its existence calls into question the long-term stability of RV measurements, and would limit the detectability of other Earths. Comparing solar data sets from different instruments over several years is mandatory if the community wants to demonstrate that RVs have the capability of detecting other Earth-like planets.

With this manuscript, we demonstrated the extreme precision of the decade of HARPS-N solar observations. We hope that the community will use these
data to develop novel methods to mitigate stellar signals in RVs, with the goal of enabling the detection of other Earths, but also to propel the use of extreme precision RVs for other science applications.

\begin{acknowledgements}

The success of the HARPS-N spectrograph and the solar telescope would not have been possible without the dedicated work of all the TNG team, in particular telescope operators Albar Garcia, Gianni Mainella, Hristo Stoev, Carmen Padilla and Daniele Carosati, that make sure that the solar telescope is running every possible, and technicians Marcos Hernandez and Hector Ventura, that fixed solar telescope issues over the years.
We note that the HARPS-N Instrument Project was partially funded through the Swiss ESA-PRODEX Programme and the HARPS-N solar telescope through two Smithsonian R\&D grants.\\

\boldfont{We thank Eric Ford for sharing the NEID data that allowed to compare the HARPS-N and NEID data in Figure 6.}

XD acknowledges the support from the European Research Council (ERC) under the European Union’s Horizon 2020 research and innovation programme (grant agreement SCORE No 851555) and from the Swiss National Science Foundation under the SPECTRE grant (No 200021\_215200).

KA acknowledges support from the Swiss National Science Foundation (SNSF) under the Postdoc Mobility grant P500PT\_230225.

AM, AAJ, BSL acknowledge funding from a UKRI Future Leader Fellowship, grant number MR/X033244/1. AM, FR acknowledges funding from a UK Science and Technology Facilities Council (STFC) small grant ST/Y002334/1. 

AS works was funded by the European Union (ERC, FIERCE, 101052347) and also supported by FCT - Funda\c c\~ao para a Ci\^encia e a Tecnologia through national funds by these grants: UIDB/04434/2020 DOI: 10.54499/UIDB/04434/2020, UIDP/04434/2020 DOI: 10.54499/UIDP/04434/2020, PTDC/FIS-AST/4862/2020, UID/04434/2025.

SA, BK and NKOS acknowledge the support from the European Research Council (ERC) under the European Union’s Horizon 2020 research and innovation programme (grant agreement SCORE No 865624).

VB acknowledges the support from the European Research Council (ERC) under the European Union's Horizon 2020 research and innovation programme (project {\sc Spice Dune}, grant agreement No 947634). 

ACC and PCZ acknowledge support from STFC consolidated grant number ST/V000861/1
and UKRI/ERC Synergy Grant EP/Z000181/1 (REVEAL).

HMC acknowledges funding from a UKRI Future Leader
Fellowship (grant numbers MR/S035214/1 and MR/Y011759/1).

CAW and JC acknowledge support from the UK Science and Technology Facilities Council (STFC, grant number ST/X00094X/1).

ZLD and AV acknowledge support from the D.17 Extreme Precision Radial Velocity Foundation Science program under NASA grant 80NSSC22K0848.
ZLD would like to thank the generous support of the MIT Presidential Fellowship, the MIT Collamore-Rogers Fellowship and to acknowledge that this material is based upon work supported by the National Science Foundation Graduate Research Fellowship under Grant No. 1745302.

RDH is funded by the UK Science and Technology Facilities Council (STFC)'s Ernest Rutherford Fellowship (grant number ST/V004735/1).

AL acknowledges support of the Swiss National Science Foundation under grant number TMSGI2\_211697

MS thanks the Belgian Federal Science Policy Office (BELSPO) for the provision of financial support in the framework of the PRODEX Programme of the European Space Agency (ESA) under contract number C4000140754.

FPE and CLO acknowledge the Swiss National Science Foundation (SNSF) for supporting research with HARPS-N through the SNSF grants nr. 140649, 152721, 166227, 184618 and 215190.\\

This work has been carried out in the framework of the National Centre for Competence in Research \emph{PlanetS} supported by the Swiss National Science Foundation (SNSF). This research used the DACE platform developed in the framework of PlanetS (\url{https://dace.unige.ch}).
The results presented in this document rely on data produced by the Institute of Environmental Physics (IUP) at the University of Bremen (https://www.iup.uni-bremen.de/eng/) and are available at https://www.iup.uni-bremen.de/gome/gomemgii.html. These data were accessed via the LASP Interactive Solar Irradiance Datacenter (LISIRD) (https://lasp.colorado.edu/lisird/).

\end{acknowledgements}

\bibliographystyle{aa}
\bibliography{dumusque_bibliography}

\begin{appendix}

\section{Major HARPS-N instrumental interventions}\label{app:intervention}

We list in Table ~\ref{tab:intervention_details} all the major HARPS-N instrumental interventions since the change of the detector in November 2012.

\begin{table}[h]
  \centering
    \caption{HARPS-N major intervention information.}
  \begin{tabular}{|c|c|}
    \hline
    \textbf{Date} & \textbf{Intervention Type} \\ \hline
    12.11.2012 & Change of HARPS-N detector \\ \hline
    05.03.2013 & warm-up \\ \hline
    17.03.2014 & focus changed \\ \hline
    18.06.2014 & warm-up \\ \hline
    17.10.2014 & warm-up \\ \hline
    07.11.2014 & TH lamp flux ratio change \\ \hline
    03.02.2015 & warm-up \\ \hline
    19.02.2015 & warm-up \\ \hline
    23.02.2015 & warm-up \\ \hline
    14.05.2015 & TH lamp flux ratio change \\ \hline
    19.05.2015 & warm-up \\ \hline    
    13.10.2015 & warm-up \\ \hline
    31.03.2016 & warm-up \\ \hline
    26.10.2016 & warm-up \\ \hline
    11.04.2017 & warm-up \\ \hline
    14.11.2017 & warm-up \\ \hline
    23.04.2018 & warm-up \\ \hline
    15.06.2018 & TH lamp flux ratio change \\ \hline
    21.10.2018 & warm-up \\ \hline
    12.03.2019 & warm-up \\ \hline
    19.02.2020 & TH lamp flux ratio change \\ \hline
    24.03.2020 & TH lamp flux ratio change \\ \hline
    27.03.2020 & TH lamp flux ratio change \\ \hline
    14.04.2020 & TH lamp flux ratio change \\ \hline
    01.06.2020 & TH lamp changed \\ \hline
    19.07.2019 & warm-up \\ \hline
    22.12.2019 & warm-up \\ \hline
    28.05.2020 & warm-up \\ \hline
    19.07.2020 & warm-up \\ \hline
    27.11.2020 & warm-up \\ \hline
    16.02.2021 & TH lamp flux ratio change \\ \hline
    12.04.2021 & warm-up \\ \hline
    25.06.2021 & warm-up \\ \hline
    10.10.2021 & cryostat replacement \\ \hline
    28.10.2021 & warm-up \\ \hline
    11.05.2022 & warm-up \\ \hline
    18.12.2023 & warm-up \\ \hline
  \end{tabular}
  \label{tab:intervention_details}
\end{table}

\section{Correction for Thorium shift due to lamp ageing}\label{app:th_ageing}

\begin{figure*}
        \center
	\includegraphics[width=6cm]{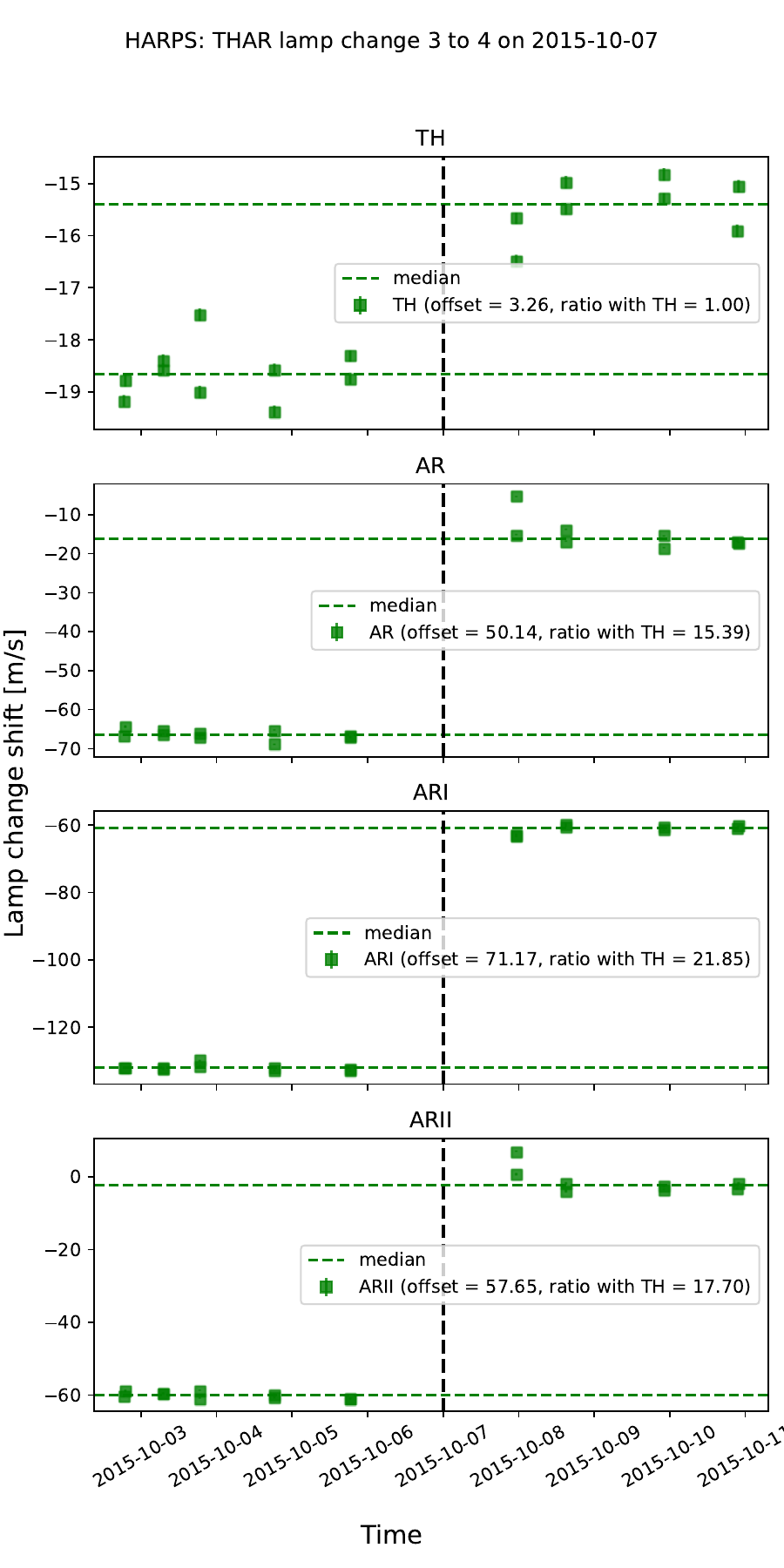}
	\includegraphics[width=6cm]{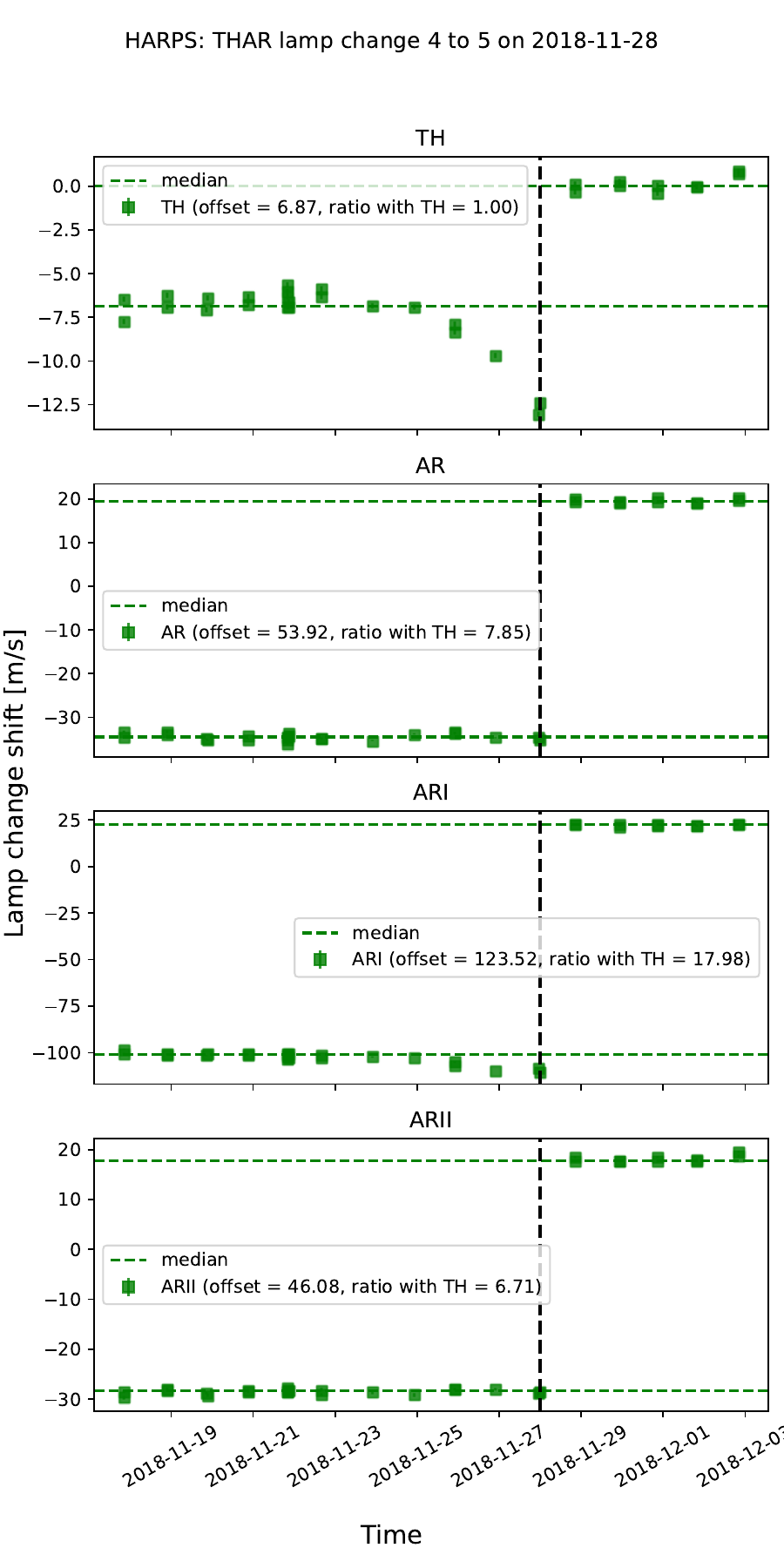}
	\includegraphics[width=6cm]{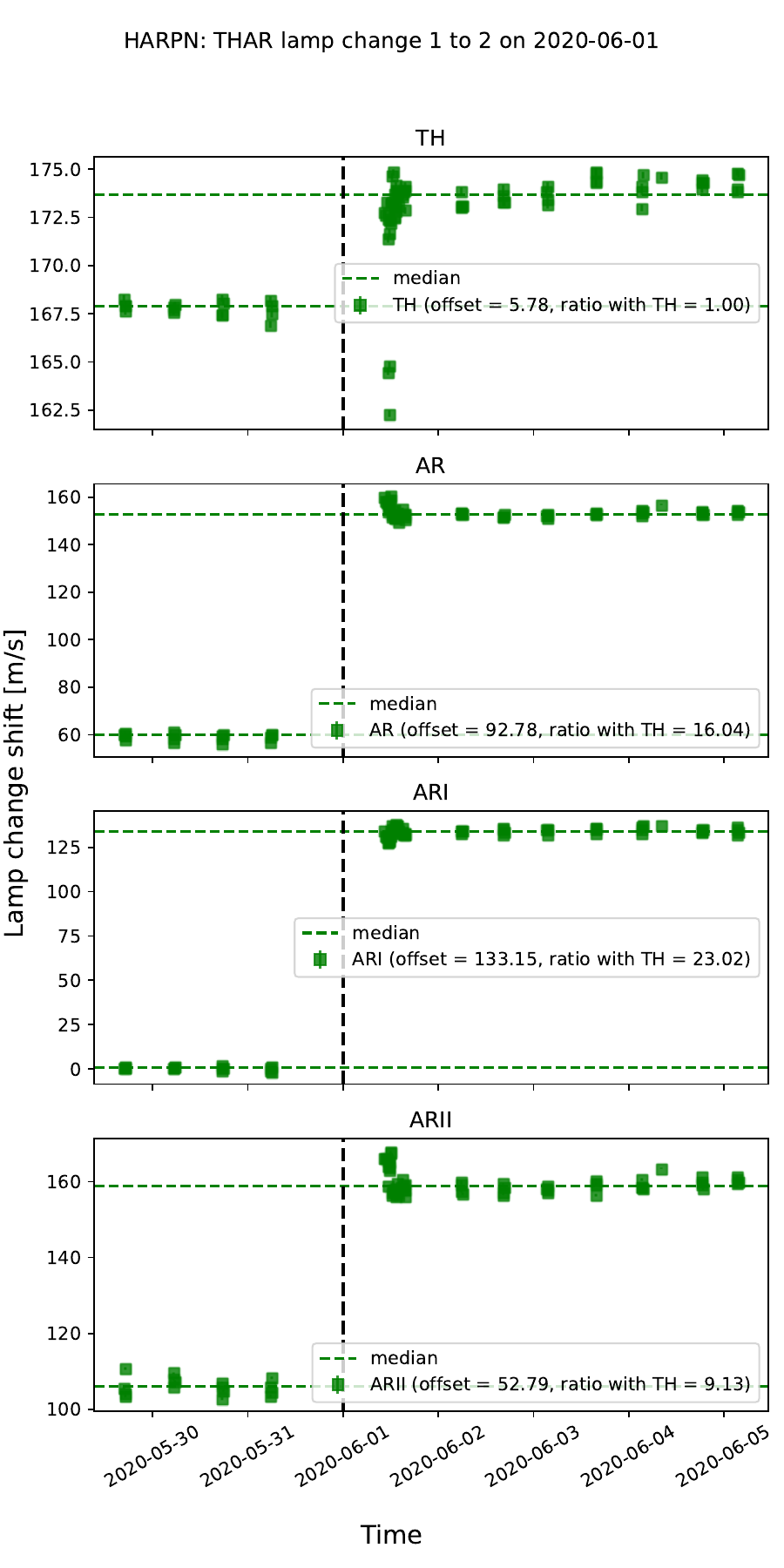}
	\caption{From top to bottom, observed offset of TH, AR, AR1 and AR2 lines at the two last change of TH-AR HC lamp on HARPS (on 2015-10-07 (\emph{left}) and on 2018-11-28 (\emph{middle})) and the only lamp change on HARPS-N on 2020-06-01 (\emph{right}). We clearly see an offset in the position of TH lines between an \emph{old} and a \emph{new} lamp, of the order of a few \ms, that goes alongside a much larger effect observed on AR lines. Among those three lamp changes, we see that the ratio between AR1 and TH lines offset is rather consistant, with values of 21.9, 18.0 and 23.0, while this is not the case for AR2 lines, and therefore also all AR lines combined.}
	\label{fig:th_offset_lamp_change}
\end{figure*}

Fig.~\ref{fig:th_offset_lamp_change} highlights the offset in velocity of TH and AR lines at three different lamp changes, one for HARPS-N and two for HARPS.
As we can see, TH and AR lines present a significant offset, with the former being smaller than the latter. Among the three lamps, we observed that for the ground state Argon lines (AR1),
the AR1 to TH offset ratio is on average 20.95. We cannot however measure the shift of AR1 lines through time and use this 20.95 value to correct for the TH drift, as
a measurement of the drift of AR1 lines will require wavelength solutions which are themselves derived from TH lines.
Therefore, when we change a lamp, we can measure the AR1 to TH offset ratio, while when measuring the drift of AR1 through time, we measure the drift difference
between AR1 and TH, $\mathrm{offset}_{AR} - \mathrm{offset}_{TH}$. Following the
development below, it is however possible to use the AR1 to TH offset ratio to measure the effective drift of TH lines:
\begin{align}
\frac{\mathrm{offset}_{AR1}}{\mathrm{offset}_{TH}} &= X\\
\mathrm{offset}_{TH} &= \frac{\mathrm{offset}_{AR1} - \mathrm{offset}_{TH}}{X} + \frac{\mathrm{offset}_{TH}}{X}\\
\mathrm{offset}_{TH} &= \frac{\mathrm{offset}_{AR1} - \mathrm{offset}_{TH}}{X-1}\\
\end{align}

\section{Thorium line curation for optimising long-term wavelength solution stability}\label{app:th_curation}

To perform a new curation of TH lines, we started from the \citet{Lovis-2007b} catalogue and only selected lines associated with TH or AR. We then \edit{cross-matched these lines} with the \citet{Redman:2014aa} catalogue and kept only lines in our original line selection that had a single line in \citet{Redman:2014aa} within 1\kms and for which the ionisation level was identical. We also replaced the \citet{Lovis-2007b} wavelengths and errors by the ones from \citet{Redman:2014aa}, as being more precise and accurate. After this selection, we were left with 4053 TH (2961 TH1,  1079 TH2 and 13 TH3) and 447 AR (125 AR1 and 322 AR2) lines. 

We then ran the Gaussian line fitting algorithm of the ESPRESSO DRS for all TH-AR lamp calibrations for more than 10 years, from November 2012\footnote{HARPS-N started its operation early 2012, but half of the CCD stopped working in September 2012, which required replacing the device and a major intervention on HARPS-N in October 2012. The ESPRESSO DRS has been optimised only to reduce data obtained with the new CCD that is still in operation.} to June 2023. The DRS does some quality checks on the Gaussian fit performed on all TH and AR lines. Lines are flagged as bad if they are too close to the edge of the orders, saturated, too weak or embedded in the noise \footnote{Maximum flux minus the average of the right and left line minima smaller than 100.}, if the line fitted full width at half maximum is not between zero and six pixels \footnote{TH lines are not resolved and the instrumental resolution is oversampled on average by a factor of 3.5.}, or if the error in pixel position is larger than one. As in \cite{Dumusque:2021aa}, we first selected the lines that were used more than 99\% of the time, to use \edit{a consistent set of lines} to perform wavelength solutions and other analyses. This criterion reduces the number of valid TH and AR lines to 3486 and 408, respectively.

These rather basic quality checks do not flag lines that \edit{show non-negligable variation} over time or that are sensitive to the overall flux of the lamp. We therefore performed additional checks to test the stability of the lines over time and reject the ones that show significant variability. First, for each TH spectrum analysed, in total 9452 spanning over 3862 days, we measured the residual of the wavelength solution computed as the difference between the tabulated wavelengths of all TH lines in our line list \citep[from][]{Redman:2014aa} and the wavelengths derived from the wavelength solution given the line fitted pixel positions. We then rejected 209 lines for which the absolute mean wavelength difference over time was more than 200\ms as those always deviate from the wavelength solutions\footnote{We converted the wavelength difference to a Doppler shift using the formula $\delta v = c\,.\,\delta\lambda/\lambda$, where $c$ is the speed of light.}. We also performed a cut in peak-to-peak RV variation of all TH lines to reject lines with too much variability. To be robust to outliers, the peak-to-peak amplitude of each line is measured as the difference between the 90th and the 10th percentile. A threshold at 200\ms rejects an additional 54 lines, reducing the total number of valid TH lines to 3223.

We then measure the velocity drift of all TH lines at each observation epoch\footnote{We note that the velocity drift of each TH line was measured by comparing the pixel position at each time with respect to a reference (timestamp of reference TH frame: 2017-11-12T15-34-18.950), then multiplying this difference by the dispersion of the spectrograph at the position of the line (the width of the pixel in wavelength) to obtain a drift in wavelength and then dividing this number by the wavelength of each line before multiplying by the speed of light to obtain a drift in velocity.}. Over more than 10 years of operation, the HARPS-N spectrograph drifted by no more than a few hundred \ms (see top panel of Fig.~\ref{fig:th_ar_drift}). To investigate whether this drift is the same everywhere across the detector, we divided the HARPS-N detector into 36 sectors, 
\edit{forming a six-by-six grid in the dispersion and cross-dispersion direction}
\footnote{pixel separation [1,683,1366,2048,2731,3413,4096] and order separation [1,12,23,35,46,57,69].}.
\edit{In each sector, we measured} the weighted average velocity drift of the respective TH lines once the median velocity drift over all lines was removed. In Fig.~\ref{fig:th_drift_detector}, we can see strong residuals that differs for different regions of the HARPS-N detector, caused by local LSF variation through time. The largest deviations, as high as 300\ms, appear in the blue and red edges of the HARPS-N wavelength range. To mitigate the effect of those LSF variations, which will be calibrated out each time a new wavelength solution is performed \edit{(i.e., on a daily basis)}, we \edit{subtracted the local median velocity drift from all TH line velocity drift measurements}. We then use those TH velocity drifts locally corrected for LSF variation effects to calculate for each line the quantities defined in Table~\ref{tab:TH_observables}.
\begin{table*}[h]
  \centering
  \begin{tabular}{llc}
    \hline \hline
    \textbf{Name} & \textbf{Description} & \textbf{Threshold investigated} \\ \hline
    $\mathrm{S}_{time}$ & Slope of the RV drift as a function of time divided by the fitted Gaussian width & [3,4,\textbf{5}] $\sigma$ \\ 
      & of the distribution for all TH lines & \\
    $\mathrm{A}_{corr}$ &Pearson correlation coefficient between the RV drift and its fitted amplitude &  [0.3,\textbf{0.5},0.7] \\ 
    $\mathrm{FR}_{corr}$ & Pearson correlation coefficient between the RV drift and the TH spectrum flux ratio\footnote{The flux ratio is the ratio of the total flux of one TH calibration with respect to the reference TH frame 2017-11-12T15-34-18.950.} & [0.3,0.5,\textbf{0.7}] \\
    $Z_{score}$ & Ratio between the RV drift MAD and median photon-noise error & [3,4,\textbf{5},] $\sigma$\\ 
    $O_{intervention}$ & Maximum offset in RV drift among all major instrument intervention divided by & [2, \textbf{3}, 4] $\sigma$\\ 
       & the median photon-noise error (see Appendix~\ref{app:intervention})  &  \\ \hline
    \end{tabular}
  \caption{Observables for TH line selection optimisation and investigated threshold to reject lines with significant variability. Values in bold font correspond to the threshold used for our optimal line selection\edit{, listed in Table~\ref{tab:th_selection}.}}
  \label{tab:TH_observables}
\end{table*}
%

\begin{figure*}
        \center
	\includegraphics[width=9cm]{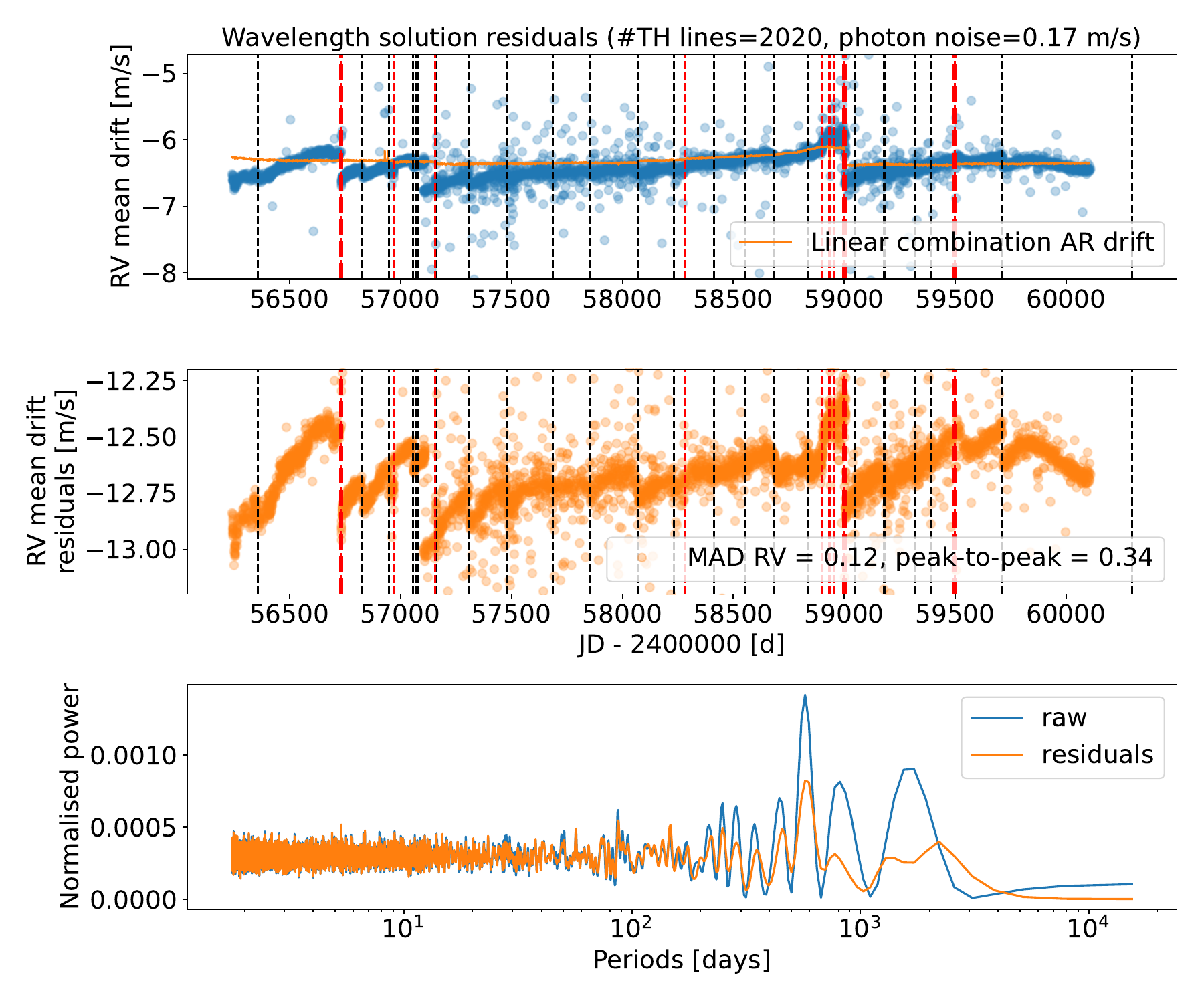}
	\includegraphics[width=9cm]{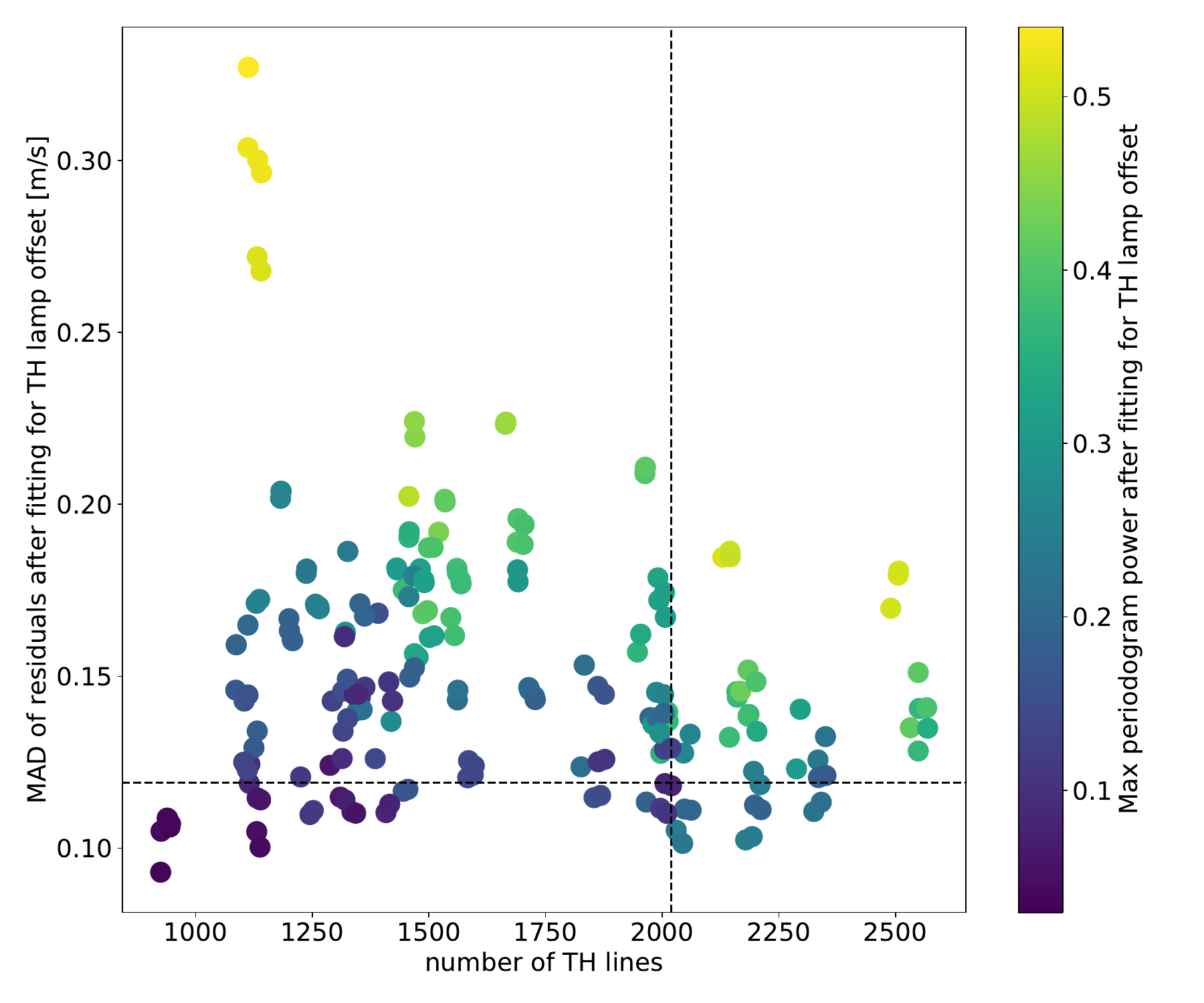}
	\caption{\emph{Left:} Wavelength solution residuals for our best selection of TH lines. A wavelength solution residual is the difference between the known wavelength of TH lines (as shown in Table~\ref{tab:th_selection}) and the wavelengths obtained from a derived wavelength solution. \emph{Top left:} Mean RV drift of wavelength solution residuals though time in blue. The orange curve corresponds to the best-fitted linear model of the AR drift shown in the middle panel of Fig.~\ref{fig:th_ar_drift} to account for the drift of TH lines with respect to lamp aging. The residuals after this curve have been removed are shown in the \emph{middle left panel}. \emph{Bottom left:} Periodogram of the mean RV drift and mean RV drift residuals. \emph{Right:} MAD of mean RV drift residuals as a function of the number of lines selected for deriving a wavelength solution. The colour coding highlights the power of the maximum peak found in the periodogram of the mean RV drift residuals. The dashed lines points to the best selection of TH line used for deriving the HARPS-N wavelength solutions used for reducing the solar data. The left panel shows the wavelength solution residuals for that selection.} 
	\label{fig:wavesol_residuals}
\end{figure*}
\begin{figure*}
        \center
	\includegraphics[width=18cm]{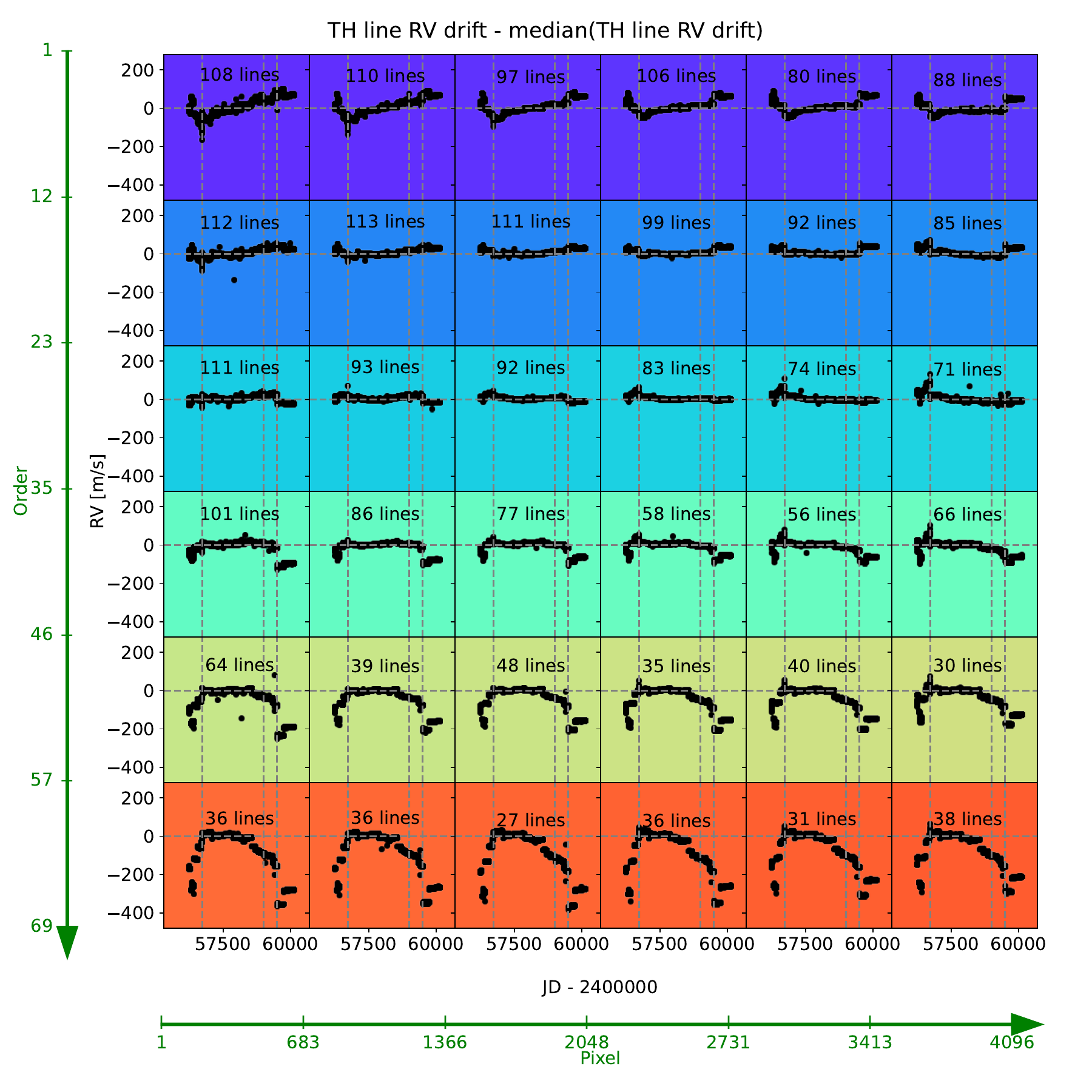}
	\caption{Drift of TH lines as a function of time for different regions on the detector. The green axis on the left and at the bottom of the plot gives the delimitation in order and pixel of each detector region, respectively. Each subplot shows the weighted average velocity drift of the TH lines falling in a specific region of the detector once the overall drift of all TH lines has been removed (shown in the top panel of Fig.~\ref{fig:th_ar_drift}). The text in each subplot gives the number of TH lines falling in each specific region. The dashed vertical lines correspond, from left to right, to a modification of the HARPS-N focus, the change of the TH-AR HC lamp used for wavelength solution, and the replacement of the detector cryostat. The blue and red edges show strong long-term deviations, likely due to significant LSF variation in those regions over time, in addition to significant jumps at the focus change and cryostat replacement.} 
	\label{fig:th_drift_detector}
\end{figure*}

Optimising the selection of TH lines used for wavelength solution is a delicate trade-off between selecting \edit{lines which are less affected by systematics} while \edit{retaining enough lines that are} well sampled over the different spectral orders to provide a precise enough wavelength solution.
To perform this optimisation, we tested the different threshold values listed in Table~\ref{tab:TH_observables} to reject unstable TH lines. For each line selection, we derived the wavelength solutions for all the 9452 HARPS-N TH-AR spectra available. For each wavelength solution, we measured the residual of the solution for all lines used and performed a weighted average of those residuals. We show the variation of this global wavelength residual over time in the top left panel of Fig.~\ref{fig:wavesol_residuals}. The panel figure reveals a significant offset at JD = 2459002 (1st June 2020), which corresponds to the change of the TH-AR lamp used for wavelength calibrations. As discussed in Sec.~\ref{sect:th_drift}, this can be mitigated by measuring the drift observed in AR1 lines (as shown in the middle panel of Fig.~\ref{fig:th_ar_drift}) and applying a scaling factor to this drift to model the drift of TH lines. We therefore performed this correction by fitting for that scaling factor, and obtained the residuals shown in the middle left panel of Fig.~\ref{fig:wavesol_residuals}. We then computed the \edit{median absolute deviation (MAD)} of the obtained residuals and its peak-to-peak amplitude defined as the difference between the 90th and the 10th percentile. We also computed the amplitude of the highest peak in the periodogram of those residuals (seen in the bottom left panel of the same figure). In the right panel of Fig.~\ref{fig:wavesol_residuals}, we show for each combination of the thresholds defined in Table~\ref{tab:TH_observables} the MAD of the wavelength residual time series as a function of the number of TH lines used, with the power of the maximum peak in the periodogram as colour scale. The combination of the smallest wavelength solution residual MAD and smallest amplitude for the highest periodogram peak provides the best selection of lines for long-term wavelength solution stability. However, reducing the number of TH lines used \edit{too drastically} will decrease the global wavelength solution precision, in addition to providing \edit{fewer} lines to the 
2D polynomial fit performed for wavelength solution. The fit will be less constrained on the edges of the detector, where flux is overall lower (lower on the edges of the orders due to the blaze response and lower in the blue edge due to detector sensitivity). This will induce spurious variation of the wavelength solution at those locations \citep[see Appendix B of][]{Dumusque:2021aa}. Taking those two arguments into consideration, we selected as best line selection the one falling at the crossing of the dashed lines in the right panel of Fig.~\ref{fig:wavesol_residuals}. \edit{This selection} presents one of the smallest wavelength solution residual MADs, the smallest maximum peak amplitude among the neighboring simulations and in total 2020 TH lines, which provide a wavelength solution precision of 0.17\ms. The good simulations with about a thousand lines reach an even smaller MAD and maximum peak amplitude, however, the precision is \edit{roughly twice as large}, $\sim0.30$m/s. 

With this optimal selection of 2020 TH lines, that can be found in Table~\ref{tab:th_selection}, we obtain for 13 years of TH calibrations a precision in wavelength solution well below 1\ms. During the first two years of HARPS-N operations until JD = 2457104 (22 March 2015), we see some strong systematics at the level of 0.4\ms that are due to a focus change and an unknown issue linked to an instrument intervention
%
%
, however, those do not impact solar observations as they started on JD = 2457218 (14 July 2015). After that date, the
only strong systematic we have is an offset of 0.3\ms on JD = 2459002 (1 June 2020), due to the change of the TH-AR HC lamp used for wavelength solutions. Even when including those systematics, the residual around the wavelength solution for the entire lifetime of HARPS-N is smaller than 0.75\ms peak-to-peak.


%

\section{Tailored selection of thorium lines for the new HARPS-N DRS}\label{app:th_line_selection}

Table~\ref{tab:th_selection} lists the information relative to the curated TH lines used in this paper to reduce HARPS-N solar data. This table only gives an overview of the information available online.
%
\begin{table}
	\footnotesize
	\caption{Thorium line list used by the ESPRESSO DRS to reduce the HARPS-N solar data published along this manuscript. Wavelengths are in the vacuum and are coming from \citet{Redman:2014aa}. This table only shows a small fraction of the data that are available in electronic format.}            
	\label{tab:th_selection}    
	\centering                         
	\begin{tabular}{lccc}       
		\hline\hline
		Wavelength [$\AA$] & Element & HARPS-N order & HARPS-N pixel \\
		\hline
3876.74453 &  TH1 &      1 &   130.890533 \\
3878.70792 &  TH1 &      1 &   292.684570 \\
3883.42445 &  TH1 &      1 &   689.175162 \\
3884.86879 &  TH1 &      1 &   812.918648 \\
3885.62491 &  TH2 &      1 &   878.063875 \\
$\cdots$ & $\cdots$ & $\cdots$ & $\cdots$\\
6891.20409 &  TH2 &     69 &  2890.945939 \\
6894.15243 &  TH1 &     69 &  3059.980018 \\
6896.13247 &  TH1 &     69 &  3175.116339 \\
6904.11434 &  TH1 &     69 &  3650.686722 \\
6910.89548 &  TH1 &     69 &  4071.171106 \\
		\hline
	\end{tabular}
\end{table}
%

\section{Instrumental drift correction optimisation}\label{app:drift_correction}

\begin{figure*}
        \center
	\includegraphics[width=16cm]{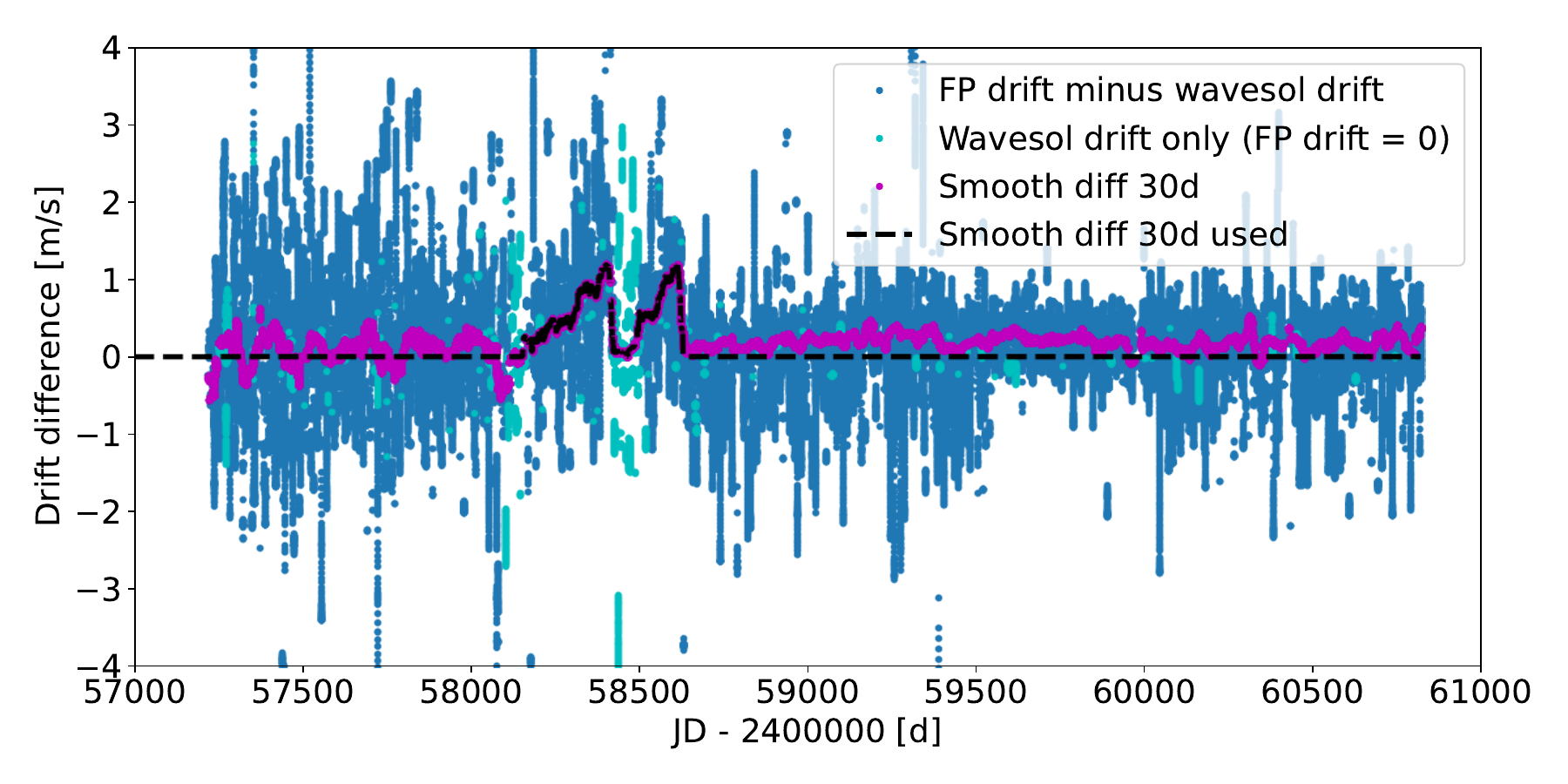}
	\caption{Difference in the spectrograph drift measurement between the information provided by simultaneous FP calibrations and a linear interpolation between daily wavelength solutions (dark blue dots). Light blue dots correspond to times when the FP was down and thus no simultaneous drift measurement was performed. The magenta dots correspond to a 30-day smoothed version of the RV difference and the black dash-line corresponds to our correction model for spurious FP drift measurements (equal to zero when the drift difference is negligible).} 
	\label{fig:drift_diff}
\end{figure*}

Between January 31st 2018 and May 25th 2019 (JD = 2458150 and 2458629), the FP was unstable, providing some spurious FP drift not tracking the instrument. To correct for those we compared the drift measured using the FP attached to solar observations to a linear interpolation between day-to-day wavelength solutions. We clearly see in Fig.~\ref{fig:drift_diff} that between January 31st 2018 and May 25th 2019, the two drift measurements diverge by up to 1\ms, with a sawtooth pattern linked to interventions on the FP calibration source. All the RVs measured for the Sun during those times when the FP was unstable were corrected by the smoothed drift difference (black curve in Fig.~\ref{fig:drift_diff}).
%

The \edit{aforementioned} issue with the FP stability led us to investigate whether the FP drift measurement used to correct for instrumental instability throughout the daily solar observation was properly capturing the systematics of HARPS-N \edit{over the duration of the entire data set}. 
The FP drift measurement photon-noise precision varied between 0.05 and 0.15\ms from the start of the solar observation\edit{s} up to July 25th 2018, then between 0.1 and 0.3\ms until the upgrade of the FP system on May 25th 2019.
\edit{The precision since then has systematically been better than 0.1\ms}
It can happen, in some cases where the instrument is very stable, that applying the FP drift correction adds noise to the data. We compared for each day three different ways of measuring and applying drift corrections: i) no drift correction, ii) using the FP and iii) linearly interpolating between daily wavelength solutions in addition to adding the high-frequency FP drift variations \footnote{We high-pass filtered the FP drift with a window of 4 hours, by removing a rolling mean over 4 hours using the python algorithm pandas.rolling("4h", min\_periods=1, center=True).mean().}. We note that adding the high-frequency FP drift variations for method iii) \edit{significantly improved} the results of this method, therefore pointing towards instrumental variations on timescales below 4 hours. 

\edit{We compared the different drift corrections on a day-by-day basis. We measured two statistics to evaluate the best method to use: the daily RV rms, and the 5-day RV rms.}
Although the choice to optimise the daily RV rms is obvious, the minimisation of the 5-day RV rms ensures that when nitrogen is refilled every week, which creates several\ms drift of the instrument, we keep the correction consistent with a smooth evolution of the spectrograph drift over days. Although we are aware that stellar activity varying on a timescale of 25 days due to solar rotation will tend to slightly influence this 5-day RV rms, this window ensures that always several days of solar observation are included, even in case of bad weather. If a drift correction method \edit{gave} the best 5-day RV rms and the corresponding daily RV rms was smaller than 0.6\ms or not 20\% larger than the smallest value, this drift correction method was selected as the optimal one for that day. Alternatively, if the best method \edit{did not improve} the 5-day RV rms by more than 20\%, then the method giving the smallest daily RV rms was selected for that day. These simple criteria automatically selected the best drift correction method for 2366 days out of the 2382 days analysed. For the remaining 16 days, we \edit{manually selected} the best drift correction method. In the end, we used for 1144, 773 and 449 days the FP drift, the wavelength solution interpolated drift plus high-frequency FP drift variations and no drift correction, respectively.

\section{Effect of blaze variation on the derived RVs} \label{app:blaze_corr}

By analysing flat-field calibrations, we realised that the blaze function was significantly varying with time, and it was corelated with the periodic warm-ups of the detector. \boldfont{Before describing the effect induced by blaze variation, we discuss the effect induced by colour variation of the spectrum due to varying observing conditions, which is well known and can help to understand how blaze variation impacts CCF-derived RVs.} 

To compute an RV, \boldfont{the ESPRESSO DRS first performs a cross correlation} 
on each spectral order individually, before summing all the individual CCFs into a global CCF. 
\boldfont{The BIS SPAN is extracted from the CCF by first computing the CCF bisector as the curve connecting the midpoints between the two wings of the CCF, and then computing the difference between the value at the top and bottom of the curve \citep[][]{Queloz-2001}.} The DRS then fits an inverted Gaussian to this global CCF in order to extract the \boldfont{RV, FWHM and CONTRAST as the mean, FWHM and normalized amplitude of the Gaussian.} As the mask used to perform cross-correlation is never a perfect match to the observed stellar spectrum, the CCF of each spectral order will present an RV bias from the stellar systemic velocity. When combining the CCF of all orders together, any relative flux variation between those orders will induce a different weight on those individual biases, thus producing a spurious RV shift. As a result, RVs are correlated with S/N as generally exposure time are fixed and a lower S/N implies more atmospheric extinction, and with it more absorption of the bluer wavelengths. The ESPRESSO DRS corrects for this colour effect, \boldfont{by modifying the flux balance of an extracted spectrum so that it matches a predefined colour balance depending on spectral type}\footnote{\boldfont{In the case of the Sun, the reference colour was extracted from HARPS-N solar observations because the light path from the solar feed to the spectrograph is different than for night-time observations, and bluer wavelengths are more absorbed due the 40-meter long fiber transporting Sun light from the solar feed to the HARPS-N calibration unit.}}. \boldfont{More precisely, the predefined colour balance corresponds to the summed flux in each spectral order of a reference spectrum, normalized to the flux in the arbitrarily defined physical order 110 (order 49 in the ESPRESSO DRS for HARPS-N). To modify the colour balance of an extracted spectrum to the one of reference, the flux in each order is summed, normalised to physical order 110, and then divided by the reference colour balance providing the colour ratio. An 8th degree polynomial is then fitted on the central wavelength of each order and the corresponding colour ratio, and finally the flux from the extracted spectrum is divided by this polynomial evaluated at all wavelengths. We note that in the process, the error bars from the extracted spectrum are not modified, as otherwise this colour correction would not have any effect when errors are propagated}.

\begin{figure*}
        \center
	\includegraphics[width=16cm]{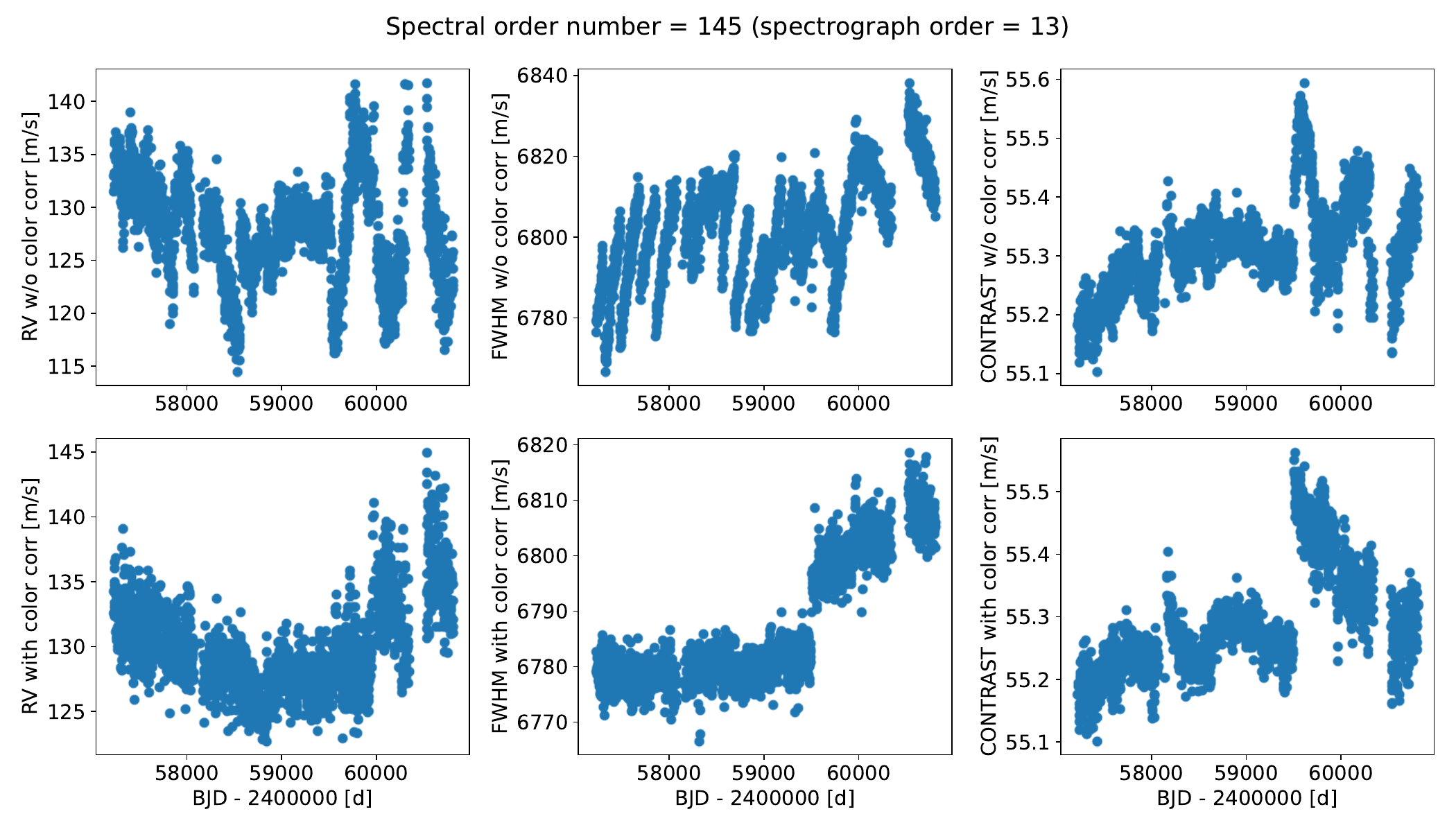}
	\includegraphics[width=16cm]{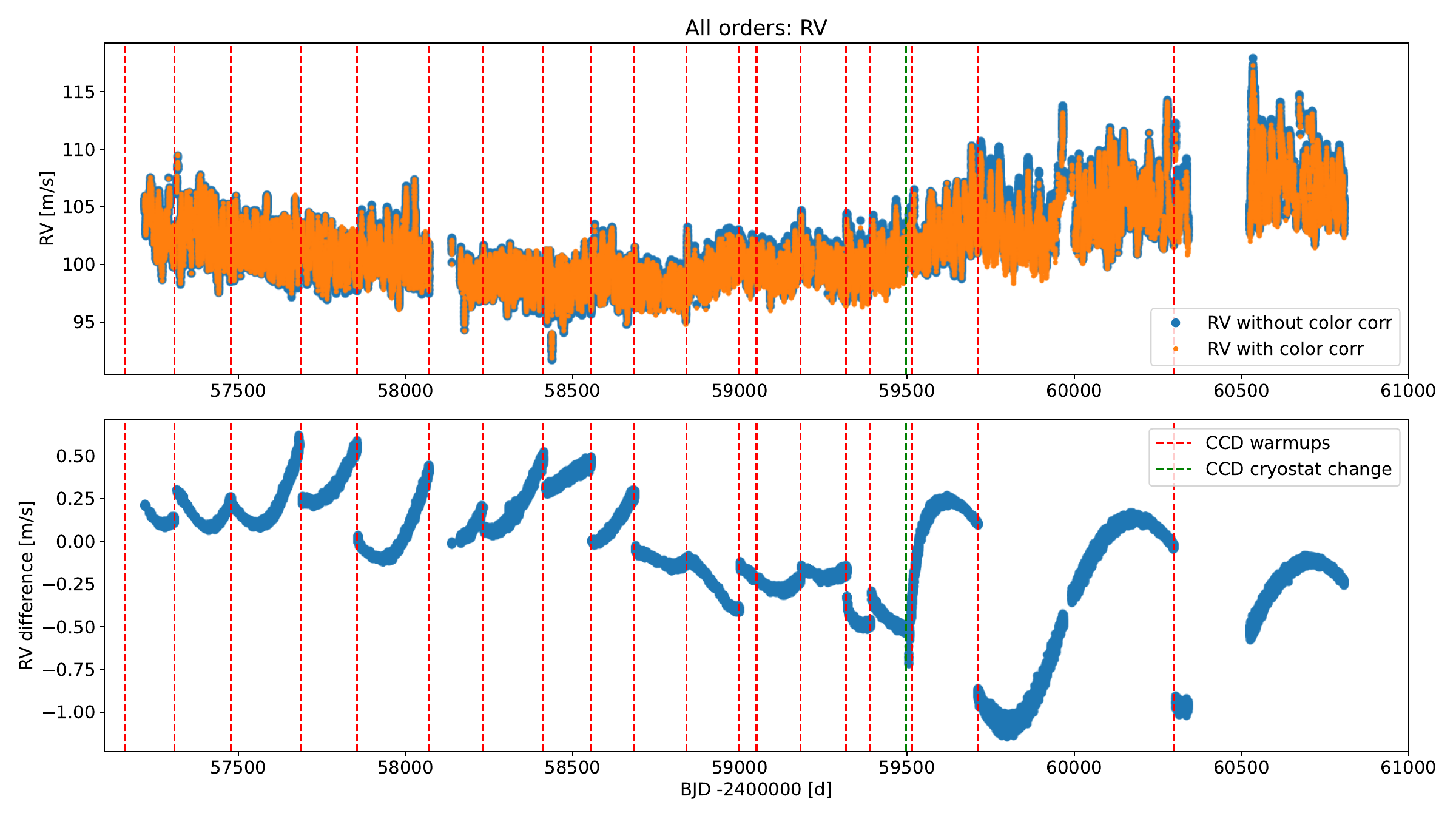}
	\caption{\emph{Top:} RV, FWHM and CONTRAST of the CCF for spectral order 145 (spectrograph order 13) as a function of time, with no correction for blaze variation on top and correction for it in the bottom (see text for details). We clearly see that the blaze variation correction significantly reduces the saw tooth pattern induced by periodic detector warm-ups. The obvious jump seen in the corrected FWHM and CONTRAST is induced by the change of the detector cryostat. \emph{Bottom:} RVs averaged across all orders with and without the correction for blaze variation. In their difference (bottom panel), we see up to 1 \ms\ jumps at each detector warm-ups shown as the red vertical dashed lines.}
	\label{fig:color_corr_order}
\end{figure*}

Similarly, a second-order effect happens if the blaze function of the instrument varies with time. In such a case, the weight given to lines appearing at different positions across a spectral order will change with time, thus inducing a spurious RV variation. To measure the effect induced by this second order systematics, we compute the CCF \boldfont{from colour-corrected extracted data (see discussion above)} with and without a correction for the blaze function variation through time. As with the colour effect discussed above, \boldfont{this correction consists in modifying the flux variation over all orders so that it corresponds to a reference blaze. In the ESPRESSO DRS, the blaze is computed from the extracted spectrum of a flat-field calibration by running a low-pass filter on each order, and normalising by the maximum value obtained for each order. To correct an extracted spectrum from blaze variation, we divide this spectrum by its corresponding blaze and multiply it by a reference blaze. Then a CCF and corresponding RV, FWHM, BIS SPAN and CONTRAST are computed as described in the beginning of the section.} We can see in the upper panel of Fig.~\ref{fig:color_corr_order}, for spectral order 145 of HARPS-N (spectrograph order 13), the difference between the RV, FWHM, and CONTRAST of the CCF before and after applying \boldfont{the correction for blaze variation. We note that the correction for colour variation discussed in the preceding paragraph was already applied to all data shown in this figure}. The saw-tooth pattern observed on the variables before correction was already shown in \citet{Meunier:2024aa} and is unsurprisingly correlated with the warm-up cycles of HARPS-N. \boldfont{Spectral order 145 is among the orders that are mostly affected by this effect, but all of them show systematics at detector warm-ups.} Correcting for this second-order effect allows us to significantly mitigate the observed systematics. In the bottom panel of Fig.~\ref{fig:color_corr_order}, we compare the RVs, averaged across all orders, with and without this second-order correction. In the RV difference, we can see significant jumps at each warm-up at the level of 1\ms. This effect due to blaze variation with time is therefore not negligible at the level of precision of the solar data, and in Sect.~\ref{sect:opti_th_drift}, we will investigate the benefit of correcting for it.

We note that the effects induced by colour variation and blaze variation are inherent to the CCF technique, where imperfect masks are used to perform cross-correlations. Template matching \citep[e.g.][]{Anglada-Escude-2012,Zechmeister:2018aa,Silva:2022aa} and line-by-line techniques \citep[][]{Dumusque:2018aa,Artigau:2022aa, Lafarga:2023aa} are not affected by such issues as they require a template built from the same stellar observations, thus removing any \boldfont{RV bias between the spectrum and the template used for RV derivation}.

\section{Description of the data release}\label{app:data_release}

The data release consists of the following data products:
\begin{itemize}
\item The extracted echelle-order spectra, corrected from the instrumental blaze, in the Solar System barycenter rest-frame. These products are called S2D spectra due to their two dimensional shape.
The first and second extensions in the FITS file contain the blaze-corrected extracted flux per pixel and corresponding error for each spectral order. 
The error corresponds to the photon-noise plus read-out noise of the detector added in quadrature, and divided by the blaze, so that the corresponding error can be directly used with the flux given in the first extension of the FITS file. The third extension corresponds to the quality flag of the pixels for each order, zero being good, and anything else being bad. Hot and bad pixels are flagged that way. Extensions four and five are the wavelength solution in the vacuum and in air, and extensions six and seven are the width of pixels in wavelength in the vacuum and in air, respectively. We note that all wavelengths are in \r{A}ngstr{\"o}m. Because of dispersion, the size of each pixel in \r{A}ngstr{\"o}m will change with wavelength, which implies that for a given order, the continuum of an S2D spectrum will show a significant slope. To get a flat continuum, the easiest way is to transform the wavelengths in a logarithmic scale.

\item The extracted merged spectra, corrected from the instrumental blaze, in the Solar System barycentric rest-frame. These products are called S1D spectra. The only extension in the FITS file includes the wavelength in the vacuum and in air for each point of the merged spectrum, its flux and the quality flag of the point, as defined in the first item above. We note that merged-1D spectra are interpolated on a grid constant in velocity space and not in wavelength space. The step between each point is 0.82\kms, equivalent to the width of a pixel in velocity space.
\item The CCFs obtained by cross-correlating the S2D spectra with the ESPRESSO G2 mask derived from solar observations, thus optimised for the Sun. The first extension in the FITS file gives the CCF measured for each echelle order, with a step of 0.82\kms, in 
addition to the photon-noise weighted average CCF over all orders. Therefore, the extension has the shape $N_{CCF} \times (N_{ord}+1)$, where $N_{CCF}$ is the number of points of the CCF and $N_{ord}$ is the number of echelle orders, 69 for HARPS-N. The second extension gives the photon noise errors, and the third one gives the quality flag of each point as defined in the first item above.
\end{itemize}
In addition to those data products, we also release the time-series of relevant quantities such as different activity indicators, the RV, the BIS SPAN, the FWHM and the CONTRAST of the CCF, but also information related to the observations (e.g. airmass, solar coordinates, S/N for different spectral orders) and the different correction applied to the data (Barycentric to heliocentric RV correction, differential extinction correction and optimised drift correction). All the information available are summarised in the README.pdf file available at this address \url{https://dace.unige.ch/downloads/open_data/dace-h4s8lp7c/files/others/README.pdf}. 

The data can be accessed using the Data Analysis Center for Exoplanet (DACE) web-platform at this address \url{https://dace.unige.ch/sun/}, or using the DACE python API (\url{https://dace.unige.ch/pythonAPI/} with the solar spectroscopy module). Bulk downloads are available on the main web page, to download all the available products (S2D, S1D and CCF), only the CCF products, or the time-series. Using the solar spectroscopy data base (go to the main web page and then follow the link to the \emph{Solar spectroscopy database}), it is possible to scan through all the released data, select specific observations by using filters (magnifying glass on the right of each column) and download the S2D, S1D or CCF products of specific observations by selecting the rows of interest (click on them), and then using one of the S2D, S1D or CCF buttons on the top right corner of the table. We note that although by default only eight columns are shown on the \emph{Solar spectroscopy database} web page, it is possible to get access to more information by clicking on the gearwheel on the top left corner of the table. All this can also be done using the DACE python API, for people who want to directly access the data within a python script. A tutorial to download the solar data and perform some basic analysis is available at this address \url{https://dace.unige.ch/pythonAPI/?tutorialId=800}. We note that regarding the time series, not exactly the same information is available from the data release and the DACE API. The differences are highlighted in the README.pdf file available at this address \url{https://dace.unige.ch/downloads/open_data/dace-h4s8lp7c/files/others/README.pdf}. \boldfont{In addition, we also uploaded the time series on the CDS VizieR database for ensuring long-term archiving}.

\end{appendix}

\end{document}